\documentclass[prb,twocolumn,showpacs,floatfix]{revtex4-1}
\usepackage{graphicx}
\usepackage{amssymb}
\usepackage{amsmath}
\usepackage{xspace}
\usepackage{url}
\usepackage{float}

\begin{document}
\newcommand{\nag}{\phantom{\dag}}
\newcommand{\comment}[1]{!!! \textbf{#1} !!!}
\title{Nonequilibrium transport through molecular junctions in the quantum regime}
\author{T. Koch$^1$, J. Loos$^2$, A. Alvermann$^3$, H. Fehske$^1$}
\affiliation{$^1$Institut f{\"u}r Physik,
            Ernst-Moritz-Arndt-Universit{\"a}t Greifswald,
             DE-17489 Greifswald, Germany\\
             $^2$Institute of Physics, Academy of Sciences of the Czech Republic, CZ-16200 Prague, Czech Republic \\
        	 $^3$Theory of Condensed Matter, Cavendish Laboratory, Cambridge CB3 0HE, United Kingdom 
}
\date{\today}
\begin{abstract}
We consider a quantum dot, affected by a local vibrational mode and contacted to
macroscopic leads, in the nonequilibrium steady-state regime. We apply a variational Lang-Firsov transformation and solve the equations of motion of the Green
functions in the Kadanoff-Baym formalism up to second order in the interaction coefficients.
The variational determination of the transformation parameter through minimization of the thermodynamic potential allows us to calculate
the electron/polaron spectral function and conductance
for adiabatic to anti-adiabatic phonon frequencies
and weak to strong electron-phonon couplings. We investigate the qualitative impact of the
quasiparticle renormalization on the inelastic electron tunneling spectroscopy signatures and
discuss the possibility of a polaron induced negative differential conductance. 
In the high-voltage regime we find that the polaron level follows the lead chemical potential to
enhance resonant transport. 
\end{abstract}
\maketitle
\section{Introduction}
Recent advances in nano-technology have made possible the creation of electronic devices with the active element being a single organic molecule. Such molecular junctions may be an alternative to semi-conductor technology in the search for further miniaturization and novel transport properties. They can be described as quantum dots, i.e., as systems of finite size coupled to macroscopic leads acting as charge reservoirs. As with metallic or semi-conducting junctions, energy level quantization determines transport. In addition, when being occupied by charge carriers, molecular quantum dots are susceptible to structural changes that may be induced by the interaction with optical phonons. As a consequence, vibrational signatures show up in the current-voltage characteristics. Moreover, they render inelastic tunneling spectroscopy (IETS) the primary experimental tool for the identification and characterization of molecular quantum dots.\cite{HM93,*Reea97,Paea00} 
 
For a thorough understanding of the underlying transport mechanisms suitable theoretical models have to be studied. The simplest one is based on a modified Fano-Anderson model where the static impurity is replaced by a single site coupled to a local phonon mode. Then the current is given by the interacting dot spectral function and the voltage bias between the noninteracting macroscopic leads.\cite{MW92a} The transport properties of the system strongly depend on the relative time scales of the electronic and phononic subsystems.\cite{GRN07} 
 
In the regime of fast electron motion and weak electron-phonon (EP) coupling standard perturbation theory applies.\cite{Caea71,MAM04,EIA09} Here, IETS signatures result from the interference of (quasi)elastic and inelastic tunneling processes.\cite{Pe88,MTU02} The calculated line shapes in the total current are found to be especially sensitive to changes in the dot-lead coupling parameter and the dot level energy.\cite{MTU03,GRN04} In general, both these quantities should be affected by conformational changes of the molecule. In the equilibrium situation, the question remains whether vibrational coupling leads to a broadening \cite{MAM04} or narrowing \cite{EIA09} of the linear conductance resonance as a function of the dot level. 
 
On the other hand, in molecular quantum dots the vibrational frequency can be larger than the kinetic energy of incident electrons. 
From the study of the Holstein molecular crystal model,\cite{Ho59a,*Ho59b} it is well-known that in this regime, strong EP interaction may heavily reduce the ``mobility" of the electrons through the formation of small polarons (electrons dressed by phonon clouds).\cite{WF97,*WF98a,*FT07,FWLB08} Consequently, for quantum dots, the formation of a local polaron is considered a possible mechanism for the observed nonlinear transport properties, such as hysteresis, negative differential conductance (NDC) and switching.\cite{GRN05,YGR07,*GNR08,LD07,AB03,*Br07,*AB09} Approaches based on the application of a Lang-Firsov transformation \cite{LF62,WJW89} to the Hamiltonian suggest that the vibrational structure of the polaron state is revealed by distinct steps in the current-voltage signal.\cite{LM02,*ZB03,MAM04,GNR06} Here, electron transport takes place via resonant tunneling through phonon sidebands.

In this paper, we investigate steady-state transport through molecular quantum dots for small-to-large dot-lead coupling and weak-to-strong EP interaction. Using the Meir-Wingreen current formula,\cite{MW92a} our main task is the determination of the interacting electronic spectral function of the quantum dot. 
%
As the background of our calculations we choose the formalism of Kadanoff-Baym,\cite{KB62} which relies on the correspondence of the nonequilibrium Green functions of complex times to the real-time response functions. Starting from the Dyson equation, the general steady-state equations for the response functions will be deduced. The solution of the latter equations will lead to a nonequilibrium spectral function which has a form analogous to the equilibrium one. The dot self-energy determining the spectral function will be calculated from the equations of motion of the Green functions up to a second order in the interaction coefficients.

Our approach is based on a variational Lang-Firsov transformation, which was developed for Holstein polarons at finite densities \cite{LHF06,*LHAF07} and recently applied to the molecular quantum dot in equilibrium.\cite{LKABF09,*KLABF10} We extend these calculations to the nonequilibrium situation and to finite temperatures, whereby the dot self-energy will be calculated self-consistently to account for the density-dependent oscillator shift. The variational parameter of the Lang-Firsov transformation is determined numerically via the minimization of the thermodynamic potential.
In this way we are able to interpolate between the self-consistent Born approximation (SCBA) \cite{HJ08,GRN04} and the small-polaron approach \cite{GNR06} previously used in the weak and strong EP coupling limits. 
We note that already in the equilibrium case, our variational calculation introduces important corrections to the corresponding spectral functions, that determine the conductance in the linear response theory. We reexamine the low-temperature equilibrium quantum dot system and analyze the occurrence of high-temperature phonon sidebands in the linear conductance. 

In the nonequilibrium situation we show the impact of the optimal polaron state on the IETS signatures mentioned above. For comparable electronic and phononic time scales, we study the crossover from coherent tunneling to sequential hopping via a transient polaron state, where the interplay of both resonant and off-resonant multiphonon processes leads to complicated electron tunneling spectra.
Recently La Magna and Deretzis \cite{LD07} applied a similar variational ansatz to an effective electron Hamiltonian and found polaron-formation-induced NDC. Considering the dependence of the current-voltage characteristics on the full spectral function, we critically discuss this effect. 

The paper is organized as follows: Sec.~\ref{SECgeneral} introduces the model Hamiltonian and describes the variational Lang-Firsov transformation. In Secs.~\ref{SECeom} and \ref{SECsteady}, a formal steady-state solution to the equations of motion is presented. In Sec.~\ref{SECselfenergy}, we derive an approximation to the polaronic self-energy that is self-consistent and depends on the variational parameter. The latter is determined from the numerical minimization of the thermodynamic potential that is deduced in Sec.~\ref{SECvarprod}. Section~\ref{SECelecpol} gives the relation between the electronic and polaronic spectral functions. In Sec.~\ref{SECcurrent}, the general current formula for arbitrary voltage is discussed and the special case of linear conductance is mentioned.
Section~\ref{SECnumres} presents our numerical results and Sec.~\ref{SECsummary} summarizes.
\section{Theory}\label{SECtheory}
%
%
\subsection{General equations}\label{SECgeneral}
Our considerations are based on the standard Hamiltonian of the single-site quantum dot model,
\begin{align}
 H  &=  (\Delta-\mu) d^\dag d^{\nag} -\; g \omega_0 d^\dag d ( b^{\dag} + b) + \omega_0  b^{\dag} b\label{EQUhamiltonianstart} \\
& + \sum_{k,a} (\varepsilon_{ka}^{\nag}-\mu) c_{ka}^{\dag} c_{ka}^{\nag} - \frac{1}{\sqrt{N}}\sum_{k, a}\left(t_{ka}d^{\dag}c_{ka}^{\nag} + t_{ka}^\ast c_{ka}^{\dag}d\right) .\nonumber
\end{align}
Here, the quantum dot is represented by the energy level $\Delta$, with the fermionic creation (destruction) operator $d^{\dag}$ ($d$). The dot is coupled to a local phonon mode $b^{(\dag)}$ of energy $\omega_0$, with $g$ being the dimensionless EP coupling strength. 
The $\varepsilon_{ka}$ (for $k=1,\dots,N$) are the energies of noninteracting electrons in the left and right lead ($a=L,R$) with the equilibrium chemical potential $\mu$. The corresponding operators $c_{ka}^{\dag}$ ($c_{ka}^{\nag}$) create (annihilate) free fermions in the $N$ lead states. The last term in Eq.~(\ref{EQUhamiltonianstart}) allows for dot-lead particle transfer.
 
We apply to the model (\ref{EQUhamiltonianstart}) a variational Lang-Firsov transformation,\cite{LF62,Feea94,LD07,LKABF09,*KLABF10} introducing two parameters $\gamma$ and $\bar\gamma$:
\begin{align}
\widetilde H&=S_2^\dag(\bar\gamma)S_1^\dag(\gamma) H S_1(\gamma)S_2(\bar\gamma)\;, \label{EQUlft}\\
S_1(\gamma)&=\exp\{\gamma g (b^{\dag}-b)d^\dag d\}\;,\\
S_2(\bar\gamma)&=\exp\{\bar\gamma g (b^{\dag}-b)\}.
\end{align}
$S_1(\gamma)$ describes the antiadiabatic limit, where the phononic time scale is much faster than the electronic time scale and the deformation of the dot adjusts instantaneously to the presence of an electron. For $\gamma=1$ it coincides with the shifttransformation of the Lang-Firsov small-polaron theory,\cite{LF62} which eliminates the second term on the right-hand side of Eq.~(\ref{EQUhamiltonianstart}) and lowers the dot level by the polaron binding energy 
\begin{align}
\varepsilon_p&=g^2\omega_0. 
\end{align}
To account for the competition between polaron localization and charge transport, an incomplete Lang-Firsov transformation with $\gamma\in[0,1]$ is used, where $\gamma$ will be determined variationally.
The second shift transformation $S_2(\bar\gamma)$ describes the regime of fast electron motion, where the quasistatic displacement of the equilibrium position of the oscillator affects transport. According to similar considerations in Ref.~\onlinecite{Feea94}, the parameter $\bar \gamma$ is fixed by the condition that the oscillator shift is stationary in the equilibrium and steady state. Then $\bar \gamma=(1-\gamma)n_d$, with the dot occupation
\begin{align}
n_d&=\langle d^{\dag}d \rangle\;,
\end{align}
where $\langle \cdots \rangle$ denotes the steady state mean value. 
%
%
%

After the transformation the Hamiltonian reads
\begin{align}
\widetilde H &=  \widetilde\eta\, d^\dag d^{\nag} -C_{d}^{\nag} (d^\dag d-n_d)  + \omega_0  b^{\dag} b + \varepsilon_p(1-\gamma)^2n_d^2 \nonumber\\
&+\sum_{k,a} \xi_{ka}^{\nag}c_{ka}^{\dag} c_{ka}^{\nag} -\sum_{k,a} \left(  C_{ka}^{\nag} d^{\dag}c_{ka}^{\nag}+ C_{ka}^\dag c_{ka}^{\dag}d\right)  \;,\label{EQUhamiltonian}
\end{align}
with 
\begin{align}
\widetilde\eta&=\Delta-\mu-\varepsilon_p\gamma(2-\gamma)-2\varepsilon_p(1-\gamma)^2n_d\;,\label{EQUdefeta}\\
 \widetilde{g}&=\gamma g\;,\quad\xi_{ka}=\varepsilon_{ka}-\mu\;,\label{EQUdefenergies}\\
C_{ka}&=\frac{t_{ka}}{\sqrt{N}}\,\mathrm{e}^{-\widetilde g (b^{\dag}-b)}\;,\quad C_d=g\omega_0(1-\gamma)(b^\dag+b)\;.\label{EQUdefinteraction}
\end{align}
Here $\widetilde\eta$ is the renormalized energy of the single dot level. $C_{ka}$ and $C_d$ are the renormalized interaction coefficients of the dot-lead transfer and the EP interaction, respectively.
Note that now the operators $d$ and $b$ represent dressed electrons (in analogy to polarons) and the shifted local oscillator.
The original electron and oscillator operators, now denoted by $\widetilde d$ and $\widetilde b$, read
\begin{align}
\widetilde d&= \mathrm{e}^{\;\widetilde g (b^{\dag}-b)}d\;,\quad \widetilde b=b+\widetilde g d^\dag d+(1-\gamma) g n_d\;.
\end{align}
We describe the application of a potential difference between the leads by adding to Eq.~(\ref{EQUhamiltonian}) the interaction with the external fields $\{U\}$ and define the voltage bias $\Phi$ accordingly:
\begin{align}
H_{\mathrm{int}}&=\sum_{a}U_a\sum_{k}c_{ka}^\dag c_{ka}^{\nag}\;,\quad\textnormal{with}\quad U_a=-\delta\mu_a \;,\\
\Phi&=(U_L-U_R)/\mathrm{e} \;,
\end{align}
where $e$ is the (negative) elementary charge.
The response of the quantum dot is given by the polaronic nonequilibrium real-time Green functions
\begin{align}
g_{dd}(t_1,t_2;U)&=-\mathrm{i}\langle \mathcal{T} d_U(t_1)d_U^\dagger(t_2)\rangle\;, \label{EQUdefresponse}\\
g_{dd}^<(t_1,t_2;U)&=\mathrm{i}\langle  d_U^\dagger(t_2)d_U(t_1)\rangle \;,\label{EQUdefresponseless}\\
g_{dd}^>(t_1,t_2;U)&=-\mathrm{i}\langle  d_U(t_1)d_U^\dagger(t_2)\rangle \;.\label{EQUdefresponsegtr}
\end{align}
Remember that $\langle \cdots \rangle$ denotes the equilibrium average with respect to $\widetilde H$, while the time dependence of the operators $d^{(\dagger)}$ is now given by $\widetilde H+H_{\mathrm{int}}$. The time ordering operator in Eq.~(\ref{EQUdefresponse}) is defined by
\begin{align}
\mathcal{T} d_U(t_1)d_U^\dagger(t_2) &= d_U(t_1)d_U^\dagger(t_2)\;,\quad \textnormal t_1-t_2>0\\
						& = -d_U^\dagger(t_2) d_U(t_1)\;,\quad \textnormal t_1-t_2<0
\end{align}
According to Kadanoff-Baym,\cite{KB62} the real-time response functions (\ref{EQUdefresponse})-(\ref{EQUdefresponsegtr}) may be deduced using the equations of motion for the nonequilibrium Green functions of the complex time variables $t=t_0-\mathrm{i}\tau$, $\tau\in[0,\beta]$, defined as
\begin{align}
G_{dd}(t_1,t_2;U,t_0)&=-\frac{\mathrm{i}}{\langle S \rangle}\langle \mathcal{T}_\tau d(t_1)d^\dagger(t_2)S\rangle\;, \label{EQUdefG}\\
G_{dd}^<(t_1,t_2;U,t_0)&=\frac{\mathrm{i}}{\langle S \rangle}\langle \mathcal{T}_\tau d^\dagger(t_2)d(t_1)S\rangle\;, \label{EQUdefGless}\\
G_{dd}^>(t_1,t_2;U,t_0)&=-\frac{\mathrm{i}}{\langle S \rangle}\langle \mathcal{T}_\tau d(t_1)d^\dagger(t_2)S\rangle\;,\label{EQUdefGgtr}
\end{align}
where the order of $t_1$ and $t_2$ is fixed in $G_{dd}^<$ and $G_{dd}^>$. The time dependence of all operators is determined by $\widetilde H$ and the external disturbance is explicit in the time-ordered exponential operator $S$:
\begin{align}\label{EQUexponential}
S&=\mathcal{T}_t\exp\left\{-\mathrm{i}\int_{t_0}^{t_0-\mathrm{i}\beta}\mathrm{d}t \;H_{\mathrm{int}}(t)\right\} \;.
\end{align}
In Eqs. (\ref{EQUdefG})-(\ref{EQUdefGgtr}) and (\ref{EQUexponential}) the operator $\mathcal{T}_\tau$ orders times according to
\begin{align}
\mathcal{T}_\tau d_U(t_1)d_U^\dagger(t_2) &= d(t_1)d^\dagger(t_2)\;,\quad \mathrm{i}(t_1-t_2)>0\\
						& = -d^\dagger(t_2) d(t_1)\;,\quad \mathrm{i}(t_1-t_2)<0
\end{align}
In the following, the Green functions of ``mixed" operators, $G_{cd}(k,a;t_1,t_2;U,t_0)$ and $g_{cd}(k,a;t_1,t_2;U)$, will be used, which are defined similar to Eqs.~(\ref{EQUdefresponse})-(\ref{EQUdefGgtr}).
The functions $g$ follow from the functions $G$ through the limiting procedure $t_0\to -\infty$.
%
\subsection{Equations of motion}\label{SECeom}
We consider the polaronic dot Green function (\ref{EQUdefG}),
%
%
where the index ``$dd$" will be omitted for the moment, and start from the Dyson equation in the matrix form
\begin{align}
\left[G^{(0)-1}(t_1,\bar t;U,t_0)-\Sigma(t_1,\bar t;U,t_0)\right]\bullet G & (\bar t,t_2;U,t_0)\label{EQUdyson}\\
&=\delta(t_1-t_2)\;.\nonumber
\end{align}
In Eq.~(\ref{EQUdyson}) the matrix multiplication ``$\bullet$" is defined by $\int_{t_0}^{t_0-\mathrm{i}\beta}\mathrm{d}\bar t\cdots$ and the $\delta$ function of complex arguments is understood with respect to this integration. With the inverse zeroth-order Green function
\begin{align}\label{EQUinverseG0}
G^{(0)-1}(t_1,t_2)=\left ( \mathrm{i}\frac{\partial}{\partial t_1} - \widetilde \eta \right ) \delta (t_1-t_2)\;,
\end{align}
Eq.~(\ref{EQUdyson}) gives for $\mathrm{i}(t_1-t_0)<\mathrm{i}(t_2-t_0)$ 
\begin{align}\label{EQUeomG1}
\Big(\mathrm{i}\dfrac{\partial}{\partial t_1} & -\widetilde\eta\Big)G^<(t_1,t_2;U,t_0)=\\
&\quad\int_{t_0}^{t_1}\mathrm{d}\bar t\;\Sigma^>(t_1,\bar t;U,t_0)G^<(\bar t,t_2;U,t_0) \nonumber\\
&+ \int_{t_1}^{t_2}\mathrm{d}\bar t\;\Sigma^<(t_1,\bar t;U,t_0)G^<(\bar t,t_2;U,t_0) \nonumber \\
&+\int_{t_2}^{t_0-\mathrm{i}\beta}\mathrm{d}\bar t\;\Sigma^<(t_1,\bar t;U,t_0)G^>(\bar t,t_2;U,t_0)\;,\nonumber
\end{align}
where the self-energy functions $\Sigma^\gtrless$ are defined analogously to $G^\gtrless$:
\begin{align}
\Sigma^>(t_1,t_2;U,t_0)=\Sigma(t_1,t_2;U,t_0),\quad\mathrm{i}(t_1-t_2)>0\;, \\
\Sigma^<(t_1,t_2;U,t_0)=\Sigma(t_1,t_2;U,t_0),\quad\mathrm{i}(t_1-t_2)<0 \;.
\end{align}
On the other hand, the matrix-transposed form of (\ref{EQUdyson}) yields
\begin{align} 
\Big(-\mathrm{i}\frac{\partial}{\partial t_2}& - \widetilde\eta\Big)G^<(t_1,t_2;U,t_0)= \label{EQUeomG2} \\
&\quad\int_{t_0}^{t_1}\mathrm{d}\bar t\;G^>(t_1,\bar t;U,t_0)\Sigma^<(\bar t,t_2;U,t_0)\nonumber\\
&+ \int_{t_1}^{t_2}\mathrm{d}\bar t\;G^<(t_1,\bar t;U,t_0)\Sigma^<(\bar t,t_2;U,t_0) \nonumber \\
&+\int_{t_2}^{t_0-\mathrm{i}\beta}\mathrm{d}\bar t\;G^<(t_1,\bar t;U,t_0)\Sigma^>(\bar t,t_2;U,t_0) \;. \nonumber
\end{align}
Similarly to Eqs.~(\ref{EQUeomG1}) and (\ref{EQUeomG2}), equations having $G^>(t_1,t_2;U,t_0)$ on the left-hand side are obtained in the case $\mathrm{i}(t_1-t_0)>\mathrm{i}(t_2-t_0)$. 
After the limiting procedure $t_0\to -\infty$, we arrive at the equations for the real-time response functions of the dot operators: 
\begin{widetext}
\begin{align}
\left ( \mathrm{i}\frac{\partial}{\partial t_1}-\widetilde\eta \right ) g^{\lessgtr}(t_1,t_2;U) &=\quad \int_{-\infty}^{t_1}\mathrm{d}\bar t\; \left [\Sigma^>(t_1,\bar t;U) -\Sigma^<(t_1;\bar t;U) \right ] g^{\lessgtr} (\bar t,t_2;U)  \label{EQUeomg1}\\
&\quad- \int_{-\infty}^{t_2}\mathrm{d}\bar t\;\; \Sigma^{\lessgtr}(t_1,\bar t;U)  \left [ g^{>} (\bar t,t_2;U) - g^{<} (\bar t,t_2;U) \right ]\;,\nonumber\\
\left ( -\mathrm{i}\frac{\partial}{\partial t_2}-\widetilde\eta \right )  g^{\lessgtr}(t_1,t_2;U)&=\quad \int_{-\infty}^{t_1}\mathrm{d}\bar t\; \left [g^>(t_1,\bar t;U) -g^<(t_1;\bar t;U) \right ] \Sigma^{\lessgtr} (\bar t,t_2;U)  \label{EQUeomg2}\\
 &\quad- \int_{-\infty}^{t_2}\mathrm{d}\bar t\;\; g^{\lessgtr}(t_1,\bar t;U)  \left [ \Sigma^{>} (\bar t,t_2;U) - \Sigma^{<} (\bar t,t_2;U) \right ]\;.\nonumber
\end{align}
The latter equations are general; up to this point no special assumptions or approximations were made. 
%
\subsection{Steady-state solution}\label{SECsteady}
Limiting ourselves to the steady-state regime, all functions of $(t_1,t_2)$ will be supposed to depend only on $t=t_1-t_2$. Then, after suitable change of the integration variables, the difference of the equations for $g^<$ in Eqs.~(\ref{EQUeomg1}) and (\ref{EQUeomg2}) gives
\begin{equation}\label{EQUfirst}
 \intop_{-\infty}^{\infty}\mathrm{d}\bar t\; \left [ g^<(\bar t;U)\Sigma^>(t-\bar t;U) -g^>(\bar t;U)\Sigma^<(t-\bar t;U) \right ] = 0\;,
\end{equation}
while the differential equation for $(g^>-g^<)$ following from Eq.~(\ref{EQUeomg1}) reads
\begin{align}
\left ( \mathrm{i} \frac{\partial}{\partial t} -\widetilde \eta \right ) \left [ g^>(t;U)-g^<(t;U) \right ] &=\quad \int_{0}^{\infty}\mathrm{d}\bar t \left [ \Sigma^>(\bar t ;U) -\Sigma ^<(\bar t ; U) \right]\left[ g^>(t-\bar t;U) - g^<(t-\bar t;U)\right ]\\
&\quad -\int_{-\infty}^{0}\mathrm{d}\bar t\left [ \Sigma^>( t-\bar t ;U) -\Sigma ^<(t-\bar t ; U) \right]\left[ g^>(\bar t;U) - g^<(\bar t;U)\right ] \;. \nonumber
\end{align} 
Using the Fourier transformations of $g^{\lessgtr}$ and $\Sigma^{\lessgtr}$ with factors according to Kadanoff-Baym,\cite{KB62} e.g. 
\begin{align}
g^{\lessgtr}(\omega;U) &= \mp\mathrm{i}\int_{-\infty}^{\infty} \mathrm{d}t\; g^{\lessgtr}(t;U)\mathrm{e}^{\mathrm{i}\omega t} \;,\label{EQUtrafoom}\\
g^{\lessgtr}(t;U) &= \mp\int_{-\infty}^{\infty} \frac{ \mathrm{d}\omega}{2\pi\mathrm{i}}\; g^{\lessgtr}(\omega;U)\mathrm{e}^{-\mathrm{i}\omega t}\;, \label{EQUtrafot}
\end{align}
the following exact equations for the steady-state are obtained:
\begin{align}
g^<(\omega;U)\Sigma^>(\omega;U) -g^>(\omega;U)\Sigma^<(\omega;U) &= 0\;,\label{EQUsteady1}\\[0.1cm]
\left[\omega-\widetilde\eta-\mathcal{P}\int_{-\infty}^{\infty}\frac{\mathrm{d}\omega^\prime}{2\pi}\;\frac{\Sigma^>(\omega^\prime;U)+\Sigma^<(\omega^\prime;U)}{\omega-\omega^\prime}\right]\left[g^>(\omega;U)+g^<(\omega;U)\right] &= \label{EQUsteady2}\\
\quad\quad\left[\Sigma^>(\omega;U)+\Sigma^<(\omega;U)\right]&\;\mathcal{P}\int_{-\infty}^{\infty}\frac{\mathrm{d}\omega^\prime}{2\pi}\;\frac{g^>(\omega^\prime;U)+g^<(\omega^\prime;U)}{\omega-\omega^\prime}\;.\nonumber
\end{align}
\end{widetext}
If we define, in analogy to the equilibrium expressions,\cite{KB62} 
\begin{align}
A(\omega;U) &= g^>(\omega;U)+g^<(\omega;U)\;,\label{EQUdefa} \\
g(z;U)&=\int \frac{\mathrm{d}\omega}{2\pi}\; \frac{A(\omega;U)}{z-\omega}\;,\label{EQUdefg}\\
\Gamma(\omega;U)&=\Sigma^>(\omega;U)+\Sigma^<(\omega;U) \;,\label{EQUdefGamma}\\
\Sigma(z;U)&=\int\frac{\mathrm{d}\omega}{2\pi}\;\frac{\Gamma(\omega;U)}{z-\omega}\;,\label{EQUdefSigma}
\end{align}
Eq.~(\ref{EQUsteady2}) takes the form
\begin{align}\label{EQUsteady3}
\left[\omega-\widetilde\eta-\mathrm{Re}\;\Sigma(\omega;U)\right]&A(\omega;U) \\
&= \Gamma(\omega;U)\;\mathrm{Re}\;g(\omega;U)\;.\nonumber
\end{align}
According to Eq.~(\ref{EQUdefa}) we can write
\begin{align}
g^<(\omega;U)&=A(\omega;U)\bar f(\omega;U)\;,\label{EQUgless}\\
g^>(\omega;U)&=A(\omega;U)(1-\bar f(\omega;U))\;,\label{EQUggtr}
\end{align}
introducing the nonequilibrium distribution $\bar f$, which follows from the steady-state equation (\ref{EQUsteady1}) and the definition (\ref{EQUdefGamma}) as
\begin{equation}\label{EQUbarf}
\bar f (\omega;U)= \frac{\Sigma^<(\omega;U)}{\Gamma(\omega;U)}\;.
\end{equation}
Looking for a solution $A(\omega;U)$ of Eq.~(\ref{EQUsteady3}), which would be equal to the equilibrium spectral function for $\{U\}\to0$, we assume (according to similar considerations in Ref. \onlinecite{KB62}) that $g(z;U)$ has the form
\begin{equation}\label{EQUgfrac}
g(z;U)=\frac{1}{z-\widetilde\eta-\Sigma(z;U)}\;.
\end{equation}
Together with Eq.~(\ref{EQUdefg}), Eq.~(\ref{EQUgfrac}) fulfils Eq.~(\ref{EQUsteady3}) identically, and the polaronic nonequilibrium spectral function becomes
\begin{equation}\label{EQUafrac}
A(\omega;U)=\frac{\Gamma(\omega;U)}{\left[\omega-\widetilde\eta-\mathcal{P}\int\frac{\mathrm{d}\omega^\prime}{2\pi}\;\frac{\Gamma(\omega^\prime;U)}{\omega-\omega^\prime}\right]^2+\left[\frac{\Gamma(\omega;U)}{2}\right]^2}\;.
\end{equation}
%
\subsection{Self-energy}\label{SECselfenergy}
We determine the polaron self-energy $\Sigma_{dd}$ from the equations of motion for the generalized Green functions of complex time, which were considered for the equilibrium case in Ref.~\onlinecite{LKABF09,*KLABF10}. In particular, the coupled equations for $G_{dd}$ and $G_{cd}$ read
\begin{align}
G_{dd}^{(0)-1}(t_1,\bar t)&\bullet G_{dd}^{\nag}(\bar t,t_2;U,t_0)= \delta(t_1-t_2) \label{EQUeomGdd}\\
& + \frac{\mathrm{i}}{\langle S \rangle}\langle \mathcal{T}_\tau C_d^{\nag}(t_1)d(t_1)d^\dag(t_2) S \rangle \nonumber\\
&+ \sum_{k,a} \frac{\mathrm{i}}{\langle S \rangle}\langle \mathcal{T}_\tau C_{ka}^{\nag}(t_1)c_{ka}^{\nag}(t_1)d^\dag(t_2) S \rangle\;,\nonumber\\
G_{cc}^{(0)-1}(k,a ;t_1&,\bar t;U)\bullet G_{cd}^{\nag}(k,a;\bar t,t_2;U,t_0) \label{EQUeomGcd}=\\
& \quad\frac{\mathrm{i}}{\langle S \rangle}\langle \mathcal{T}_\tau C_{ka}^{\dag}(t_1)d(t_1)d^\dag(t_2) S \rangle \;,\nonumber
\end{align}
where, in analogy to Eq.~(\ref{EQUinverseG0}), 
\begin{equation}\label{EQUinverseGcc}
G_{cc}^{(0)-1}(k,a;t_1,t_2;U)=\left(\mathrm{i}\frac{\partial}{\partial t_1} -\xi_{ka}^{\nag} -U_a^{\nag} \right) \delta (t_1-t_2)\;.
\end{equation}
To deduce the functional differential equations for the self-energy $\Sigma_{dd}=G_{dd}^{(0)-1}-G_{dd}^{-1}\,$, in addition to the physical fields $\{U\}$, we introduce the fictitious fields $\{V\}$ by adding to $H_{\mathrm{int}}$ (cf. Refs. \onlinecite{KB62,Sc66,LKABF09,*KLABF10})
\begin{equation}
\sum_{k,a}\Big[ V_{ka}(t)C_{ka}^{\nag}(t) +\bar V_{ka}(t)C_{ka}^{\dag}(t) \Big]+ V_{d}(t)C_{d}(t)\;.
\end{equation}
In the same way as in Ref.~\onlinecite{LKABF09,*KLABF10}, the averages on the right-hand side of (\ref{EQUeomGdd}) and (\ref{EQUeomGcd}) are expressed by means of the functional derivatives of Green functions with respect to $\{V\}$. The resulting functional differential equation for $\Sigma_{dd}$ is solved by iteration to the second order in the interaction coefficients defined in Eq.~(\ref{EQUdefinteraction}). The correlation functions of the interaction coefficients are evaluated supposing independent Einstein oscillators. Letting then $\{V\}\to 0$, the following self-consistent result is obtained:
\begin{align}\label{EQUSigma2}
&\Sigma_{dd}(t_1,t_2;U,t_0)  =  \Sigma_{dd}^{(1)}(t_1,t_2;U,t_0) \\
	&\quad\quad+  \; \left[g\omega_0(1-\gamma)\right]^2G_{dd}(t_1,t_2;U,t_0) F_3(t_1,t_2)\;.\nonumber
\end{align}
The result of the first iteration step,
\begin{align}\label{EQUSigma1}
&\Sigma_{dd}^{(1)}(t_1,t_2;U,t_0)  =  \sum_{k,a} |\langle C_{ka}^{\nag}\rangle|^2 G_{cc}^{(0)}(k,a;t_1,t_2;U) \\
	&\quad\quad\quad +  \sum_{k,a} |\langle C_{ka}^{\nag}\rangle|^2 G_{cc}^{(0)}(k,a;t_1,t_2;U)F_1(t_1,t_2)\;,\nonumber
\end{align}
is independent of $G_{dd}$. The quasiequilibrium nonperturbed Green functions of the leads read
\begin{align}
&\hspace{-0.15cm}G_{cc}^{(0)<}(k,a;t_1,t_2;U)=\mathrm{i}\mathrm{e}^{-\mathrm{i}\xi_{ka}(t_1-t_2)} f(\xi_{ka}+U_a)\;, \;\label{EQUGcc0lessgtr}\\
&\hspace{-0.15cm}G_{cc}^{(0)>}(k,a;t_1,t_2;U)=-\mathrm{i}\mathrm{e}^{-\mathrm{i}\xi_{ka}(t_1-t_2)}[1- f(\xi_{ka}+U_a) ]\;,\nonumber
\end{align}
with $f(x)=(\mathrm{e}^{\beta x}+1)^{-1}$. The functions $F_1$ and $F_3$ are given by $F_1^<$ and $F_3^<$ for $\mathrm{i}(t_1-t_2)<0$, and by $F_1^>$ and $F_3^>$  for $\mathrm{i}(t_1-t_2)>0$, respectively:
\begin{align}
F_1^{\gtrless}(t_1,t_2) &=  \exp\Big\{\widetilde g^2 \Big[(n_B(\omega_0)+1)\mathrm{e}^{\mp\mathrm{i}\omega_0(t_1-t_2)} \label{EQUF1}\\
 &\quad+ n_B(\omega_0)\mathrm{e}^{\pm\mathrm{i}\omega_0(t_1-t_2)}\Big]\Big\}-1\;, \nonumber\\
F_3^{\gtrless}(t_1,t_2) &= (n_B(\omega_0)+1)\mathrm{e}^{\mp\mathrm{i}\omega_0(t_1-t_2)}\label{EQUF3}\\
& \quad+ n_B(\omega_0)\mathrm{e}^{\pm\mathrm{i}\omega_0(t_1-t_2)}\;,\nonumber
\end{align}
with $n_{B}(x)=(\mathrm{e}^{\beta x}-1)^{-1}$. In Eq.~(\ref{EQUSigma2}), we perform the limit $t_0\to-\infty$ and the continuation of the complex time variables to real times, while keeping the condition $\mathrm{i}(t_1-t_2)<0$ for $\Sigma_{dd}^<$ and $\mathrm{i}(t_1-t_2)>0$ for $\Sigma_{dd}^>$. We arrive at
\begin{align}
\Sigma_{dd}^{\lessgtr}  (t_1,t_2;U)&= \Sigma_{dd}^{(1)\lessgtr}(t;U)\label{EQUSigmalessgtr} \\
&\quad+\; [(1-\gamma)g\omega_0]^2\; g_{dd}^{\lessgtr}(t_1,t_2;U) \nonumber\\
&\quad\times \Big [(n_B(\omega_0)+1)\mathrm{e}^{\pm\mathrm{i}\omega_0(t_1-t_2)} \nonumber\\
&\quad+ n_B(\omega_0)\mathrm{e}^{\mp\mathrm{i}\omega_0(t_1-t_2)} \Big ]\;,\nonumber\\
\Sigma_{dd}^{(1)\lessgtr}  (t_1,t_2;U)&=\sum_{k,a}|\langle C_{ka}\rangle |^2 \; g_{cc}^{(0)\lessgtr}(k,a;t_1,t_2;U) \;  \label{EQUSigma1lessgtr}\\
&\quad\times\Big\{I_0(\kappa) + \sum_{s\ge 1} I_s(\kappa) 2\sinh(s\theta)\nonumber\\
&\quad\times\Big [ (n_B(s\omega_0)+1)\mathrm{e}^{\pm\mathrm{i}s\omega_0(t_1-t_2)} \nonumber\\
&\quad+ n_B(s\omega_0)\mathrm{e}^{\mp\mathrm{i}s\omega_0(t_1-t_2)} \Big ] \Big \}\;,\nonumber
\end{align}
where
\begin{align}
\theta&= \frac{1}{2}\beta\omega_0\;,\quad \kappa=\frac{\widetilde g^2}{\sinh \theta}\;,\\
I_s(\kappa)&=\sum_{m=0}^{\infty}\frac{1}{m!(s+m)!}\left(\frac{\kappa}{2}\right)^{s+2m}\;,\label{EQUdefthetakappa}
\end{align}
and
\begin{align}
g_{dd}^< (t_1,t_2;U)&= -\int\frac{\mathrm{d}\omega}{2\pi\mathrm{i}}\; A(\omega;U) \bar f(\omega;U)\;\mathrm{e}^{-\mathrm{i}\omega (t_1-t_2)}\label{EQUtrafogddgtr}\;,\\
g_{dd}^> (t_1,t_2;U)&= \int\frac{\mathrm{d}\omega}{2\pi\mathrm{i}} \;A(\omega;U) \left [1-\bar f(\omega;U)\right]\mathrm{e}^{-\mathrm{i}\omega (t_1-t_2)}\label{EQUtrafogddless} \;.
\end{align}
Now we insert $|\langle C_{ka} \rangle |^2=(|t_{ka}|^2/N)\exp\{-\widetilde g^2\coth\theta\}$ in Eq.~(\ref{EQUSigma1lessgtr}) and go from the $k$-summation to the integration over the lead states with the help of the density of states of lead $a$:
\begin{align}
\frac{1}{N} \sum_{k,a} |t_{ka}|^2 \cdots &\to \sum_a \int_{-\infty}^{\infty}\mathrm{d}\omega \;|t_a(\omega)|^2 \varrho_a(\omega)\cdots\;,\\
\varrho_a(\omega)&=\frac{1}{N}\sum_k\delta(\omega-\varepsilon_{ka})\;.
\end{align}
We then Fourier transform Eq.~(\ref{EQUSigmalessgtr}) according to Eq.~(\ref{EQUtrafoom}) and, after evaluating the resulting delta functions, obtain
\begin{widetext}
\begin{align}
\Sigma_{dd}^{<}(\omega;U)&= \Sigma_{dd}^{(1)<}(\omega;U)\label{EQUSigmalessfourier} \\
&\quad +[(1-\gamma)g\omega_0]^2 \Big [ A(\omega-\omega_0;U)\bar f(\omega-\omega_0;U)n_B(\omega_0)+ A(\omega+\omega_0;U)\bar f(\omega+\omega_0;U) (n_B(\omega_0)+1) \Big]\;,\nonumber\\[0.2cm]
\Sigma_{dd}^{(1)<}(\omega;U)&= \mathrm{e}^{-\widetilde g^2\coth\theta} \sum_a \Big \{ I_0(\kappa) \Gamma^{(0)}_a(\omega+\mu)f(\omega+U_a)+ \sum_{s\ge 1} I_{s}(\kappa) 2 \sinh (s \theta)\label{EQUSigmalessfourier1}\\
&\quad\times \Big [ n_B(\omega_0) \Gamma^{(0)}_a(\omega-s\omega_0+\mu)f(\omega-s\omega_0+U_a)+ (n_B(\omega_0)+1)  \Gamma^{(0)}_a(\omega+s\omega_0+\mu)f(\omega+s\omega_0+U_a)  \Big ] \Big \} \;, \nonumber\\[0.2cm]
\Gamma^{(0)}_a(\omega)&=2\pi |t_a(\omega)|^2\varrho_a(\omega).
\end{align}
The function $\Sigma_{dd}^{<}(\omega;U)$ can be understood as a generalized in-scattering function of polaron-like quasiparticles at the dot.\cite{Da95_2} The second line in Eq.~(\ref{EQUSigmalessfourier1}) accounts for multiple-phonon emission and, if $T>0$, absorption processes.
After some algebraic manipulations of the Bose- and Fermi-functions, the first-order self-energy (\ref{EQUSigmalessfourier1}) may be written in the following form:
\begin{align}
\Sigma_{dd}^{(1)<}&(\omega;U)= \Gamma_{L}^{(1)}(\omega;U)f(\omega+U_L) + \Gamma_{R}^{(1)}(\omega;U)f(\omega+U_R)\;, \label{EQUSigmaddless1}\\[0.2cm]
\Gamma_{a}^{(1)}&(\omega;U)=\mathrm{e}^{-\widetilde g^2\coth\theta} \Big \{\; I_{0}(\kappa)  \Gamma^{(0)}_a(\omega+\mu) + \sum_{s\ge 1} I_{s}(\kappa)2\sinh(s\theta)\\
& \times \Big[  \Gamma^{(0)}_a(\omega+\mu-s\omega_0)\Big(n_B(s\omega_0)+1-f(\omega+U_a-s\omega_0)\Big)+ \Gamma^{(0)}_a(\omega+\mu+s\omega_0)\Big (n_B(s\omega_0)+f(\omega+U_a+s\omega_0) \Big)\Big] \Big \}\;.\nonumber
\end{align}
Because $\Sigma_{dd}^{>}(\omega;U)$ results from interchanging $n_B\leftrightarrow (n_B+1)$, $f\leftrightarrow (1-f)$ and $\bar f \leftrightarrow (1-\bar f)$ in Eqs.~(\ref{EQUSigmalessfourier})-(\ref{EQUSigmaddless1}), Eq.~(\ref{EQUdefGamma}) gives
\begin{align}\label{EQUGammaresult}
\Gamma(\omega;U) &= \Gamma^{(1)}(\omega;U) \\
&\quad+ [(1-\gamma)g\omega_0]^2\Big[ A(\omega-\omega_0;U)\Big( n_B(\omega_0) + 1-\bar f(\omega-\omega_0;U) \Big )+ A(\omega+\omega_0;U)\Big( n_B(\omega_0) + \bar f(\omega+\omega_0;U) \Big ) \Big ]\;,\nonumber\\[0.1cm]
\Gamma^{(1)}(\omega;U) &= \Gamma_{L}^{(1)}(\omega;U) + \Gamma_{R}^{(1)}(\omega;U)\;.
\end{align}
\end{widetext}
From Eq.~(\ref{EQUGammaresult}), the spectral function follows using Eq.~(\ref{EQUafrac}). 
For any parameter $\gamma<1$, the spectral function $A$ and distribution $\bar f$ have to be determined self-consistently. Furthermore, because the renormalized dot level defined in Eq.~(\ref{EQUdefeta}) depends on the dot occupation $n_d$, the latter has to fulfill the self-consistency condition 
\begin{equation}\label{EQUnselfcons}
n_d=\int_{-\infty}^{\infty}\frac{\mathrm{d}\omega}{2\pi}\;\bar f(\omega;U) A(\omega;U)\;.
\end{equation}
We note that for $\gamma=0$, our results are equivalent to the SCBA.\cite{HJ08} 
For $\gamma=1$ no self-consistency condition has to be fulfilled, as $\Sigma^{\phantom{(1)}}_{dd}=\Sigma^{(1)}_{dd}$ and $\widetilde\eta$ is independent of $n_d$.
%
%
%
\subsection{Variational procedure}\label{SECvarprod}
To determine the variational parameter $\gamma$, we minimize the thermodynamic potential $\Omega$, which is given by the partition function $Q$ as
\begin{equation}
\Omega=-\frac{1}{\beta}\ln Q\;.
\end{equation}
We assume the leads to be macroscopic objects which are negligibly influenced by the states of the dot. Accordingly, the contributions of the leads to $\Omega$ and to the mean energy $\langle \widetilde H \rangle$ give only additive constants. 
Since the electronic degrees of freedom of the dot are coupled to the oscillator ones by the second term on the right-hand side of Eq.~(\ref{EQUhamiltonian}), a decoupling approximation will be used to determine the electronic part of the thermodynamic potential.

As a consequence of the equation of motion, the following identity holds:
\begin{align}\label{EQUidentity}
&\left ( \mathrm{i}\frac{\partial}{\partial t_1}- \mathrm{i}\frac{\partial}{\partial t_2} \right) d^\dag(t_2)d(t_1)\Big|_{t_2=t_1}
=\\
&\quad\quad\quad\quad\widetilde\eta d^\dag(t_1)d(t_1)-C_{d} d^\dag(t_1)d(t_1)+H^\prime(t_1)\;.\nonumber
\end{align}
Here $H^\prime$ represents the part of the Hamiltonian (\ref{EQUhamiltonian}) that depends on the operators $d^{\dag}$, $d$. As an approximation, we neglect the second term on the right-hand side of Eq.~(\ref{EQUidentity}) and in $H^{\prime}$. Taking the statistical averages on both sides of Eq.~(\ref{EQUidentity}), remembering that
\begin{equation}
\langle d^\dag(t_2)d(t_1) \rangle = -\mathrm{i}g_{dd}^<(t_1,t_2;U)
\end{equation}
and using Eq.~(\ref{EQUtrafogddgtr}),
\begin{align}\label{EQUHprime}
\langle H^\prime \rangle&= \int\frac{\mathrm{d}\omega}{2\pi}\;(2\omega-\widetilde\eta)\;A(\omega;U)\bar f(\omega;U)
\end{align}
is obtained. To determine the corresponding electronic part of the thermodynamic potential, $\Omega^\prime$, we consider the canonical ensemble given by the Hamiltonian $H^\prime_\lambda=H_0+V_\lambda$, where $H_0=\widetilde\eta d^\dag d$ and $V_\lambda$ represents the interaction part of the Hamiltonian in (\ref{EQUhamiltonian}) with coefficients $\lambda C_{ka}$ and $\lambda C_{d}$, for $\lambda\in[0,1]$. Applying the result (\ref{EQUHprime}) gives
\begin{align}\label{EQUVlambda}
\langle V_\lambda \rangle_\lambda&= 2\int\frac{\mathrm{d}\omega}{2\pi}\;(\omega-\widetilde\eta)\;A_\lambda(\omega;U)\bar f(\omega;U)\;.
\end{align}
Here $\langle \cdots \rangle_{\lambda}$ denotes the dependence of the statistical average on $\lambda$ and the indices $\lambda$ on the right-hand side of Eq.~(\ref{EQUVlambda}) refer to the interaction coefficients in $H^\prime_\lambda$.
We use the well-known general relations \cite{KB62,FW71} for the determination of $\Omega^\prime$, namely
\begin{align}
\Omega^\prime&=\Omega^\prime(\lambda=1)=-\frac{1}{\beta}\ln Q(\lambda=1)\;,\\
\ln Q(\lambda=1)&=\ln Q(\lambda=0)-\beta\int_0^{1}\mathrm{d}\lambda\;\frac{1}{\lambda}\langle V_\lambda \rangle_\lambda\;,\label{EQUOmega1}
\end{align}
where
\begin{align}
\ln Q(\lambda=0) &= \ln (1+\mathrm{e}^{-\widetilde\eta\beta})\;.
\end{align}
To make the integration in Eq.~(\ref{EQUOmega1}) feasible, the general procedure leading to the thermodynamic potential outlined above will be carried out using the solution for the dot response in the first iteration step, described in the preceding section. In particular, the spectral function $A_\lambda(\omega;U)$ is determined according to Eq.~(\ref{EQUafrac}), using $\Gamma_{\lambda}^{(1)}(\omega;U)$, which is proportional to $\lambda^2$: $\Gamma_{\lambda}^{(1)}(\omega;U)=\lambda^2\Gamma^{(1)}(\omega;U)$. Similarly, $\bar f(\omega;U)$ is determined by Eq.~(\ref{EQUbarf}) using $\Sigma_{dd}^{(1)<}$ and $\Gamma^{(1)}$ on the right-hand side. Note however, that $\widetilde \eta$ will be determined from the electron density $n_d$ corresponding to the complete self-energy $\Sigma_{dd}^{\lessgtr}(\omega;U)$.

To complete the function $\Omega$ which is to be varied with respect to $\gamma$, we have to take into account the renormalization of the oscillator energy given in the first line of Eq.~(\ref{EQUhamiltonian}). We finally obtain that
\begin{widetext}
\begin{align}\label{EQUthermpot}
\Omega&=-\frac{1}{\beta}\ln (1+\mathrm{e}^{-\widetilde\eta\beta})+\varepsilon_p(1-\gamma)^2n_d^2 +\; \int_{0}^{1} \frac{\mathrm{d}\lambda}{\lambda}\int_{-\infty}^{+\infty}\frac{\mathrm{d}\omega}{\pi} 
\dfrac{(\omega-\widetilde\eta)\bar f^{(1)}(\omega;U)\;\; \lambda^2 \Gamma^{(1)}(\omega;U)}{\left[\omega-\widetilde\eta-\lambda^2 \mathcal{P}\int\frac{\mathrm{d}\omega^\prime}{2\pi}\;\frac{\Gamma^{(1)}(\omega^\prime;U)}{\omega-\omega^\prime}\right]^2 + \left[\lambda^2\frac{ \Gamma^{(1)}(\omega;U)}{2}\right]^2} \\\
&=-\frac{1}{\beta}\ln (1+\mathrm{e}^{-\widetilde\eta\beta}) +\varepsilon_p(1-\gamma)^2n_d^2 -\int\frac{\mathrm{d}\omega}{\pi}\,\bar f^{(1)}(\omega)\Bigg\{\; \frac{\widetilde\eta-\omega}{|\widetilde\eta-\omega|}  +\, \arctan\left(\frac{\omega-\widetilde \eta-\mathcal{P}\int\frac{\mathrm{d}\omega^\prime}{2\pi}\;\frac{\Gamma^{(1)}(\omega^\prime)}{\omega-\omega^\prime}}{\Gamma^{(1)}(\omega)/2}\right)\Bigg\}\nonumber\;.
\end{align}
\end{widetext}
The parameter $\gamma$ resulting from the variation of Eq.~(\ref{EQUthermpot}) is used to determine $\Sigma_{dd}^{\lessgtr}(\omega;U)$ according to Eq.~(\ref{EQUSigmalessfourier}).
The self-energy functions obtained in this way give the distribution function $\bar f(\omega;U)$ and the spectral function $A(\omega;U)$ according to Eqs.~(\ref{EQUbarf}) and (\ref{EQUafrac}), respectively.
%
%
\subsection{Relation between electronic and polaronic functions}\label{SECelecpol}

In the previous sections, the functions $A(\omega;U)$ and $ g_{dd}^<(\omega;U)$ in polaron representation were deduced. Because the current through the quantum dot will be given by the corresponding electronic functions $\widetilde A(\omega;U)$ and $\widetilde g_{dd}^<(\omega;U)$, we have to find a relation between these quantities. We start by decoupling the fermionic and bosonic degrees of freedom in the electronic dot Green function of complex times:
\begin{align}\label{EQUDefGelec}
\widetilde G_{dd} & (t_1,t_2;U,t_0)=-\frac{\mathrm{i}}{\langle S \rangle}\langle \mathcal{T}_\tau\widetilde d(t_1) \widetilde d^\dag(t_2)S\rangle\\
&\approx G_{dd}(t_1,t_2;U,t_0)\langle \mathcal{T}_\tau e^{\widetilde g(b^\dag-b)(t_1)}e^{-\widetilde g(b^\dag-b)(t_2)}\rangle\;. \nonumber
\end{align}
Assuming an independent Einstein oscillator, we find
\begin{align}
\langle \mathcal{T}_\tau & e^{\widetilde g(b^\dag-b)(t_1)}  e^{-\widetilde g(b^\dag-b)(t_2)}\rangle=\mathrm{e}^{-\widetilde g^2\coth\theta}\Big\{ I_{0}(\kappa) \\
	&+ \sum_{s\ge 1} I_{s}(\kappa) \left( \mathrm{e}^{s\theta}\mathrm{e}^{\pm\mathrm{i}s\omega_0(t_1-t_2)} + \mathrm{e}^{-s\theta}\mathrm{e}^{\mp\mathrm{i}s\omega_0(t_1-t_2)} \right ) \Big\}\;, \nonumber
\end{align}
where the upper signs correspond to $\mathrm{i}(t_1-t_2)>0$ and the lower ones to $\mathrm{i}(t_1-t_2)<0$. Going from the complex time variables to the real ones, the following relation between $\widetilde g_{dd}^{\lessgtr}(\omega;U)$ and $g_{dd}^{\lessgtr}(\omega;U)$ is obtained:
\begin{align}
\widetilde g_{dd}^{\lessgtr}(\omega;U)&=\mathrm{e}^{-\widetilde g^2\coth\theta}\Big\{ I_{0}(\kappa)g_{dd}^{\lessgtr}(\omega;U) \\
&\quad + \sum_{s\ge 1} I_{s}(\kappa) \Big( \mathrm{e}^{s\theta}g_{dd}^{\lessgtr}(\omega\pm s\omega_0;U) \nonumber \\
&\quad + \mathrm{e}^{-s\theta}g_{dd}^{\lessgtr}(\omega\mp s\omega_0;U) \Big ) \Big\}\;.\nonumber
\end{align}
With the identities
\begin{align}\label{EQUidentities2}
\mathrm{e}^{s\theta}&=2\sinh(s\theta)[1+n_B(s\omega_0)]\;,\\
\mathrm{e}^{-s\theta}&=2\sinh(s\theta)n_B(s\omega_0)\;,
\end{align}
the electronic function $\widetilde g_{dd}^{\lessgtr}(\omega;U)$ may be expanded as
\begin{align}\label{EQUelectronicg}
\widetilde g_{dd}^{\lessgtr} & (\omega;U)=\mathrm{e}^{-\widetilde g^2\coth\theta}\Big\{ I_{0}(\kappa)g_{dd}^{\lessgtr}(\omega;U)\\
&+ \sum_{s\ge 1} I_{s}(\kappa)2\sinh(s\theta)\Big( [1+n_B(s\omega_0)]g_{dd}^{\lessgtr}(\omega\pm s\omega_0;U) \nonumber\\
& + n_B(s\omega_0)g_{dd}^{\lessgtr}(\omega\mp s\omega_0;U)  \Big ) \Big\}\;.\nonumber
\end{align}
Considering Eqs.~(\ref{EQUgless}) and (\ref{EQUggtr}), the electronic spectral function is obtained in terms of the polaronic one as
\begin{align}\label{EQUelectronic}
\widetilde A & (\omega;U)=\widetilde g_{dd}^{<}(\omega;U)+\widetilde g_{dd}^{>}(\omega;U)=\\[0.1cm]
&\quad\quad\mathrm{e}^{-\widetilde g^2\coth\theta}\Big\{ I_{0}(\kappa)A(\omega;U) + \sum_{s\ge 1} I_{s}(\kappa)2\sinh(s\theta) \nonumber \\
&\quad\times\Big( \left[n_B(s\omega_0)+\bar f(\omega+s\omega_0;U)\right]A(\omega+ s\omega_0;U) \nonumber \\
&\quad+ \left[n_B(s\omega_0)+1-\bar f(\omega-s\omega_0;U)\right]A(\omega- s\omega_0;U)\Big ) \Big\}\;.\nonumber
\end{align}
%
\subsection{Current}\label{SECcurrent}
The operator of the electron current from lead $a$ to the dot reads
\begin{equation}\label{EQUcurrentop}
\hat J_a=\frac{\mathrm{i}\mathrm{e}}{\sqrt{N}} \sum_k \left [ t_{ka} \widetilde d^\dag c_{ka}^{\nag} - t_{ka}^\ast c_{ka}^\dag \widetilde d \right ]\;.
\end{equation}
To calculate the mean value $J_a=\langle \hat J_a \rangle$, the following connection of the expectation values to the real-time Green functions is used:
\begin{align}
\mathrm{i}\langle \widetilde d^\dag c_{ka}^{\nag} \rangle &= \widetilde g_{cd}^<(k,a;t_1,t_1;U)\label{EQUmeandc} \\
&= \int_{-\infty}^{\infty}\frac{\mathrm{d}\omega}{2\pi}\; \widetilde g_{cd}^<(k,a;\omega;U)\;, \nonumber\\
\mathrm{i}\langle c_{ka}^\dag \widetilde d\rangle & = \widetilde g_{dc}^<(k,a;t_1,t_1;U)\label{EQUmeancd}\\
&= -\int_{-\infty}^{\infty}\frac{\mathrm{d}\omega}{2\pi}\; [\widetilde g_{cd}^<(k,a;\omega;U)]^\ast \nonumber\,.
\end{align}

We start from the nonequilibrium Green function of the complex time variables for the electron operators, namely
\begin{equation}
\widetilde G_{cd}(k,a;t_1,t_2;U,t_0) = -\frac{\mathrm{i}}{\langle S\rangle} \langle \mathcal{T}_\tau c_{ka}^{\nag}(t_1) \widetilde d^\dag(t_2) S \rangle \;,
\end{equation}
where $S$ is given by Eq.~(\ref{EQUexponential}). From the commutators with the Hamiltonian in the electron representation, the equation of motion is obtained:
\begin{align}
\left ( \mathrm{i}\frac{\partial}{\partial t_1} - \xi_{ka} - U_a \right ) \widetilde G_{cd} & (k,a;t_1,t_2;U,t_0) =\label{EQUcomplexeom1}\\
& -\frac{ t_{ka}^\ast}{\sqrt{N}} \widetilde G_{dd} (t_1,t_2;U,t_0)\;.\nonumber
\end{align}
Equation~(\ref{EQUcomplexeom1}) can be rewritten as
\begin{align}
\widetilde G_{cd} & (k,a;t_1,t_2;U,t_0)= \\
&-\frac{ t_{ka}^\ast }{\sqrt{N}}\int_{t_0}^{t_0-\mathrm{i}\beta}\mathrm{d}\bar t\;G_{cc}^{(0)}(k,a;t_1,\bar t;U)\widetilde G_{dd} (\bar t,t_2;U,t_0)\;.\nonumber
\end{align}
Performing the limit $t_0\to -\infty$ while keeping $\mathrm{i} (t_1-t_2)<0$, the following equation for the real-time response functions is obtained:
\begin{align}
-\frac{\sqrt{N}}{t_{ka}^\ast} \widetilde g_{cd}^< & (k,a;t_1,t_2;U)= \label{EQUrealeom1}\\
&\quad \int_{-\infty}^{t_1}\mathrm{d}\bar t\; g_{cc}^{(0)>}(k,a;t_1,\bar t;U) \widetilde g_{dd}^<(\bar t,t_2;U)\nonumber\\
&+ \int_{t_1}^{\infty}\mathrm{d}\bar t\; g_{cc}^{(0)<}(k,a;t_1,\bar t;U) \widetilde g_{dd}^<(\bar t,t_2;U)\nonumber\\
&- \int_{t_2}^{\infty}\mathrm{d}\bar t\; g_{cc}^{(0)<}(k,a;t_1,\bar t;U) \widetilde g_{dd}^<(\bar t,t_2;U) \nonumber\\
&- \int_{-\infty}^{t_2}\mathrm{d}\bar t\; g_{cc}^{(0)<}(k,a;t_1,\bar t;U) \widetilde g_{dd}^>(\bar t,t_2;U)\nonumber\;,
\end{align}
where the quasiequilibrium functions of the noninteracting leads, $g_{cc}^{(0)\lessgtr}$, coincide with the expressions (\ref{EQUGcc0lessgtr}), with $(t_1-t_2)$ real.
Based on Eq.~(\ref{EQUrealeom1}), the formal manipulations presented in the Appendix, which are analogous to the considerations made in Ref.~\onlinecite{MW92a}, finally lead to the following formula for the electron current from the lead $a$ to the dot:
\begin{align}
J_a&= \frac{\mathrm{e}}{N} \sum_k |t_{ka}|^2 \int_{-\infty}^{\infty}\mathrm{d}\omega\; \delta(\omega-\xi_{ka})\label{EQUcurrenta}\\
&\quad\times\left \{ f(\xi_{ka}+U_a)  \widetilde A(\omega;U) -\widetilde g_{dd}^{<}(\omega;U)  \right \}\;\nonumber\\
&= \mathrm{e}\int_{-\infty}^{\infty}\frac{\mathrm{d}\omega}{2\pi}\; \Gamma_a^{(0)}(\omega+\mu)\nonumber\\
& \quad\times\left \{ f(\omega+U_a)  \widetilde A(\omega;U) -\widetilde g_{dd}^{<}(\omega;U)  \right \}\;,\nonumber
\end{align}
where the electronic functions $\widetilde g_{dd}^{<}(\omega;U)$, $\widetilde A(\omega;U)$ are given by Eqs.~(\ref{EQUelectronicg}) and (\ref{EQUelectronic}), respectively. 
Since $J_L=-J_R$ in steady state, the current formula acquires the well-known form\cite{MW92a}
\begin{align}\label{EQUcurrent}
J&= \frac{1}{2}(J_L-J_R)\\
&=\frac{\mathrm{e}}{2} \int_{-\infty}^{\infty}\frac{\mathrm{d}\omega}{2\pi}\;\Gamma^{(0)}(\omega+\mu)\left[ f_L(\omega) - f_R(\omega) \right ] \widetilde A(\omega;U)\;,\nonumber
\end{align}
with $f_a(\omega)=f(\omega+U_a)$. In Eq.~(\ref{EQUcurrent}), identical leads are assumed, so that $\Gamma^{(0)}(\omega)\equiv\Gamma_L^{(0)}(\omega)=\Gamma_R^{(0)}(\omega)$. As a check of our numerics, we find indeed that the condition $J_L=-J_R$ holds, as expected for the SCBA. 
For vanishing voltage bias $\Phi\to 0$, we can express the current as $J= - L \Phi$, where the linear conductance 
\begin{align}
L=\lim_{\Phi\to0}\{- J/\Phi\}
\end{align}
results from Eq.~(\ref{EQUcurrent}) as 
\begin{align}\label{EQUcoeff}
L&=\frac{\mathrm{e}^2}{2}\,\int_{-\infty}^{\infty}\frac{\mathrm{d}\omega}{2\pi}\; \Gamma^{(0)}(\omega+\mu) \,[-f'(\omega)]\,\widetilde A(\omega)\\
&=\frac{\mathrm{e}^2}{2}\beta\,\int_{-\infty}^{\infty}\frac{\mathrm{d}\omega}{2\pi}\; \Gamma^{(0)}(\omega+\mu) \,f(\omega)(1-f(\omega))\,\widetilde A(\omega)\;\nonumber
\end{align} 
and the electronic spectral function is now calculated in equilibrium.

\section{Numerical results}\label{SECnumres}
As stated above, the spectral function, dot occupation and $\gamma$ have to be evaluated self-consistently. We do this in a two step manner: ($\mathrm{i}$) for fixed $\gamma$ and a starting value $n_d$ in Eq.~(\ref{EQUdefeta}) we calculate $\Sigma^{(1)}_{dd}(\omega)$, $\Gamma^{(1)}(\omega)$. The corresponding $A^{(1)}(\omega)$ and $\bar f^{(1)}(\omega)$ are inserted for $A$ and $\bar f$ in the right-hand side of Eqs. (\ref{EQUSigmalessfourier}) and (\ref{EQUGammaresult}). All functions are then iterated until convergence, which is signalled by
\begin{equation}
\max_{\omega} \left\{ |A_{i+1}(\omega;U)-A_i(\omega;U)| \right \} < \delta\;,
\end{equation}
with $\delta$ being a predefined tolerance. In analogy to the occurrence of multiple stable solutions in the mean-field ansatz of Galperin {\it et al.},\cite{GRN05} for strong EP coupling or high voltages, several roots of Eq.~(\ref{EQUnselfcons}) may exist. We choose the root that minimizes the thermodynamic potential. ($\mathrm{ii}$) We do this for all parameters $\gamma$ to find the global minimum of $\Omega(\gamma,n_d(\gamma))$. The corresponding parameter will be referred to as $\gamma_{\mathrm{min}}$.

In the following numerical calculations, we suppose identical leads and work in the wide-band limit, so that $\Gamma^{(0)}(\omega)=\Gamma^{(0)}$ is energy independent.

The equilibrium state, as well as the transport properties of molecular junctions crucially depend on the time scales of the electronic and phononic subsystem. While the lifetime of an electron on the dot is given by the dot-lead coupling parameter, $\tau_{\mathrm{el}}\propto 1/\Gamma^{(0)}$,\cite{Da95_2} the phononic time scale is  given by the phonon energy $\tau_{\mathrm{ph}}\propto 1/\omega_0$. The ratio $\Gamma^{(0)}/\omega_0$ determines which subsystem is the faster one.
Moreover, one should compare the polaron formation time $\tau_{\mathrm{pol}}\propto 1/\varepsilon_p$ to the electron lifetime. If the latter is long enough, i.e. if the ratio $\varepsilon_p/\Gamma^{(0)}$ is large, a transient polaron can form at the dot. The parameter $g^2$ will yield the mean number of phonons it contains.  

\subsection{Equilibrium situation, low temperature}
We first consider the equilibrium low-temperature limit with $\mu_L=\mu_R=\mu_{\mathrm{eq}}=0$ and $T=0.01$. Before we study the physically more interesting regime of equal electronic and phononic time scales, we analyze the two limiting cases $\Gamma^{(0)}\gg\omega_0$ and $\Gamma^{(0)}\ll\omega_0$. In the following, $\omega_0=1$ fixes the energy unit. 
%
%
%
\subsubsection{Limiting cases}\label{SEClimits}
\begin{figure}[t]
\begin{center}
	\hspace{-0.3cm}\includegraphics[width=0.86\linewidth]{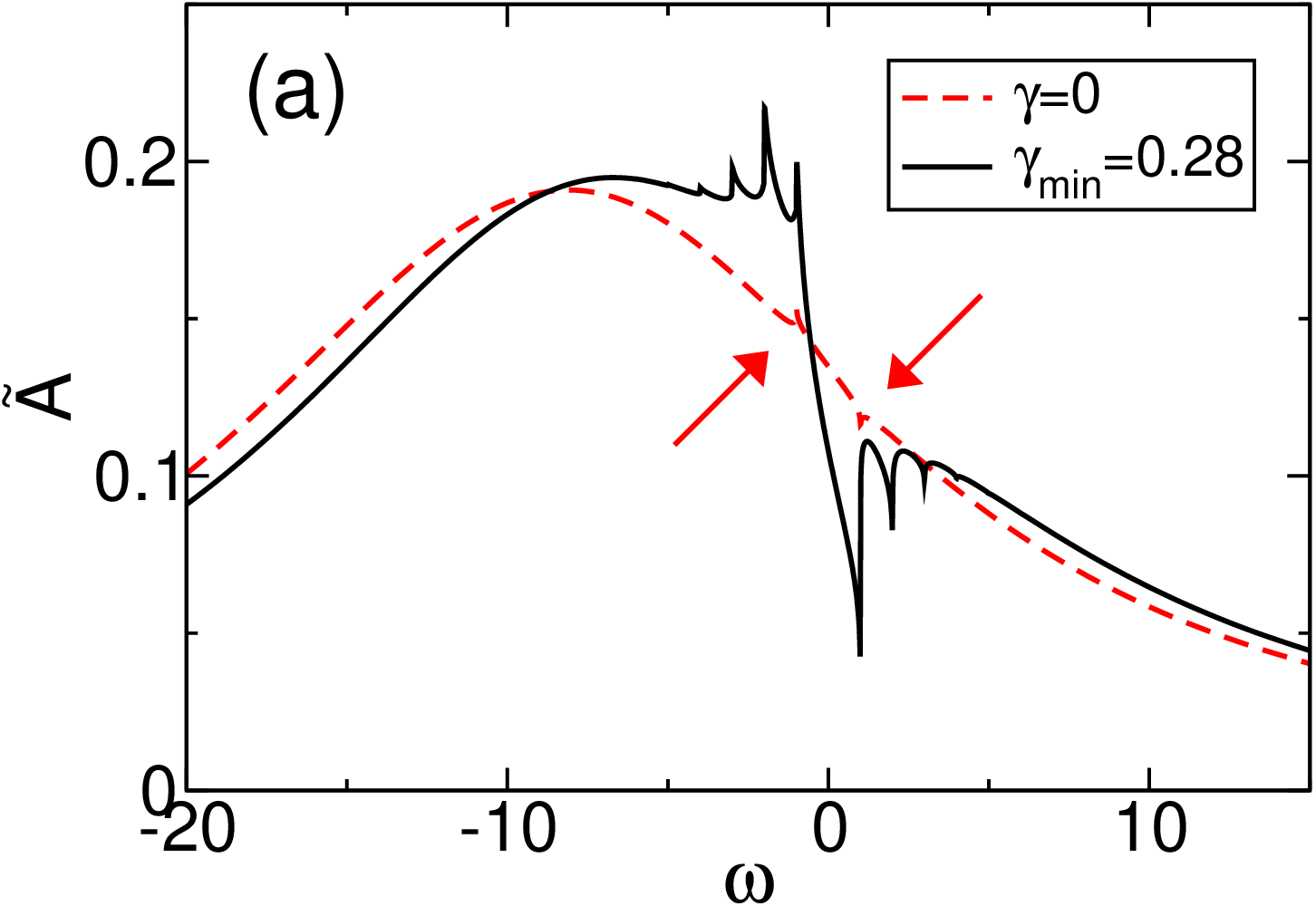}\\[0.2cm]
    	\hspace{-0.1cm}\includegraphics[width=0.85\linewidth]{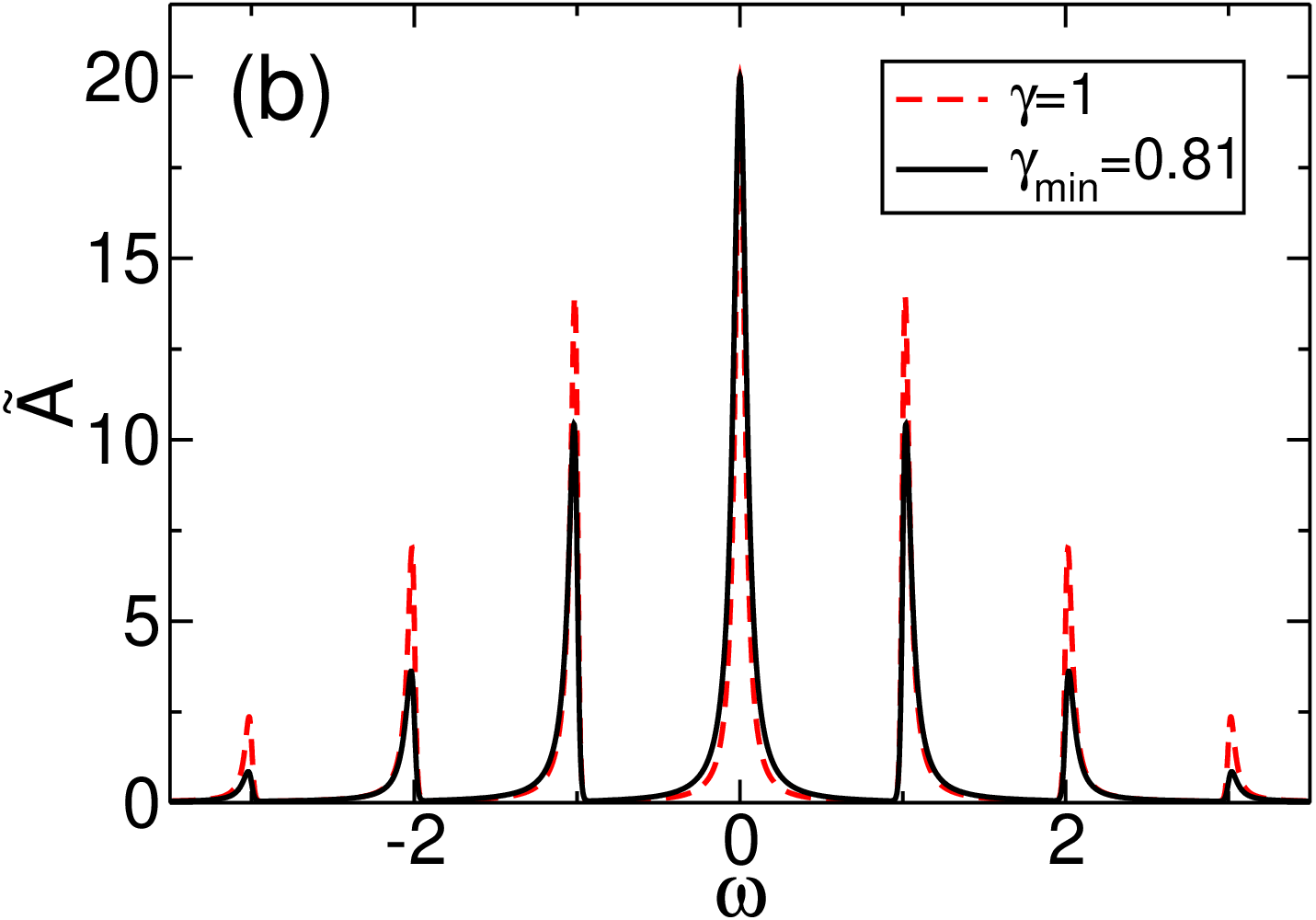}\\[0.2cm]
    	\hspace{-0.0cm}\includegraphics[width=0.9\linewidth]{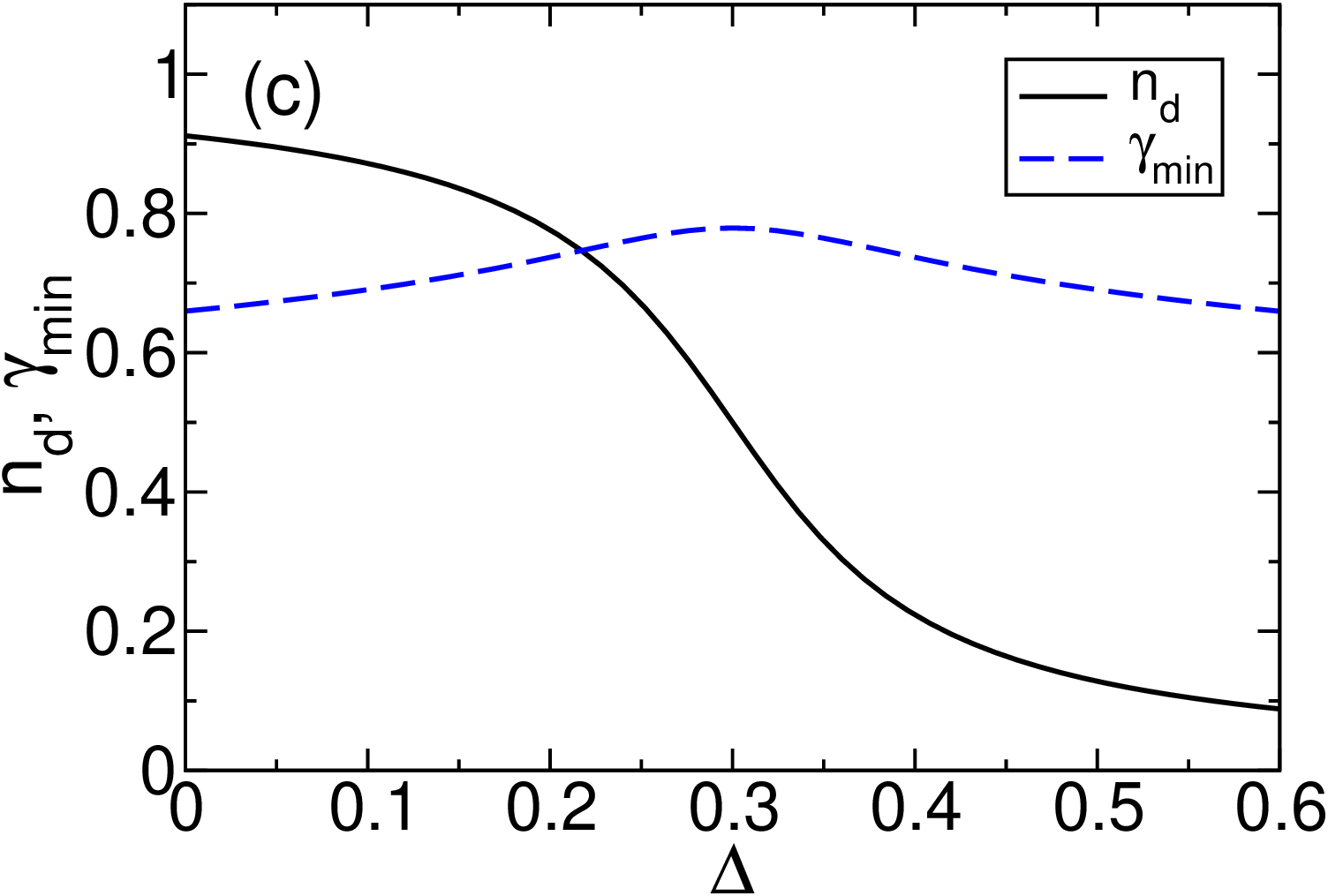}
\end{center}
\caption{For model parameters $T=0.01$, $\mu=0$ and $\Phi=0$. Panel (a): Electronic spectral functions for $\Gamma^{(0)}=10$, $\varepsilon_p=5$, $\Delta=0$ with $\gamma=0$ and $\gamma_{\mathrm{min}}=0.28$, respectively. Arrows mark the phononic features for $\gamma=0$. Panel (b): Electronic spectral functions for $\Gamma^{(0)}=0.1$, $\varepsilon_p=1$, $\Delta=1$ with $\gamma=1$ and $\gamma_{\mathrm{min}}=0.81$, respectively. Panel (c): Dot occupation and variational parameter as functions of the bare dot level $\Delta$ for $\Gamma^{(0)}=0.1$ and $\varepsilon_p=0.3$.}
\label{fig1}
\end{figure}
In the adiabatic case $\Gamma^{(0)}\gg\omega_0$, the dot deformation adjusts quasistatically to the average electronic occupation. For small EP coupling, standard perturbation approaches are applicable and the expansion of the self-energy to second order leads to the Born-approximation (BA). On a higher level, the SCBA \cite{HJ08} provides a partial resummation of the perturbation series by replacing the zero-order Green-function in the BA self-energy with the full Green function in a self-consistent way. As was mentioned above, our result (\ref{EQUSigma2}) reduces to the SCBA for $\gamma\to0$. 

Figure~\ref{fig1}(a) shows the electronic spectral function of the adiabatic quantum dot system with $\Delta=0$ and $\varepsilon_p=5$. We compare the SCBA result ($\gamma=0$) to the result of the variational calculation, yielding $\gamma_{\mathrm{min}}=0.28$. The SCBA spectrum consists of a single band, whose width is given by $\Gamma^{(0)}$. Due to the mean-field shift $\propto n_d=0.7$, the renormalized dot level lies beneath the Fermi level of the leads (at $\omega=0$) and the dot acts as a tunneling well. Because of the short residence time of electrons, the effects of inelastic scattering at the dot are small. At $\omega=-\omega_0$ ($\omega=+\omega_0$) we find a small peak (dip) in $\widetilde A$ (see arrows) due to narrow logarithmic singularities in the denominator of Eq.~(\ref{EQUafrac}).\cite{EIA09}

The variational calculation introduces several corrections to the spectrum. 
The finite $\gamma_{\mathrm{min}}$ reduces the effective mean-field coupling, i.e. the last term in the polaron shift (\ref{EQUdefeta}). Because it is not fully compensated by the $n_d$-independent contribution to Eq.~(\ref{EQUdefeta}), the overall band shifts upward. In addition, situated at integer multiples of $\omega_0$ from the lead chemical potential, several inelastic resonances form overlapping phononic sidebands. Because $\widetilde A(\omega=0)$ is lowered, transport through the dot remains coherent, but with a slightly reduced tunneling amplitude.

In the strong coupling, antiadiabatic case $\Gamma^{(0)}\ll\omega_0$, the electron occupies the dot long enough to loose coherence and interact with the phonons. Several approaches \cite{WJW89,LM02,*ZB03,GNR06} handle this regime by applying a complete Lang-Firsov transformation ($\gamma=1$) \cite{LF62} to the Hamiltonian, which gives the exact solution for the isolated molecule or when the finite occupation of the leads is neglected.\cite{Mah00} Consequently, $\gamma_{\mathrm{min}}$ can be considered a measure of the small polaron character of the dot state.

\begin{figure*}[t]
\begin{center}
\includegraphics[width=0.32\linewidth]{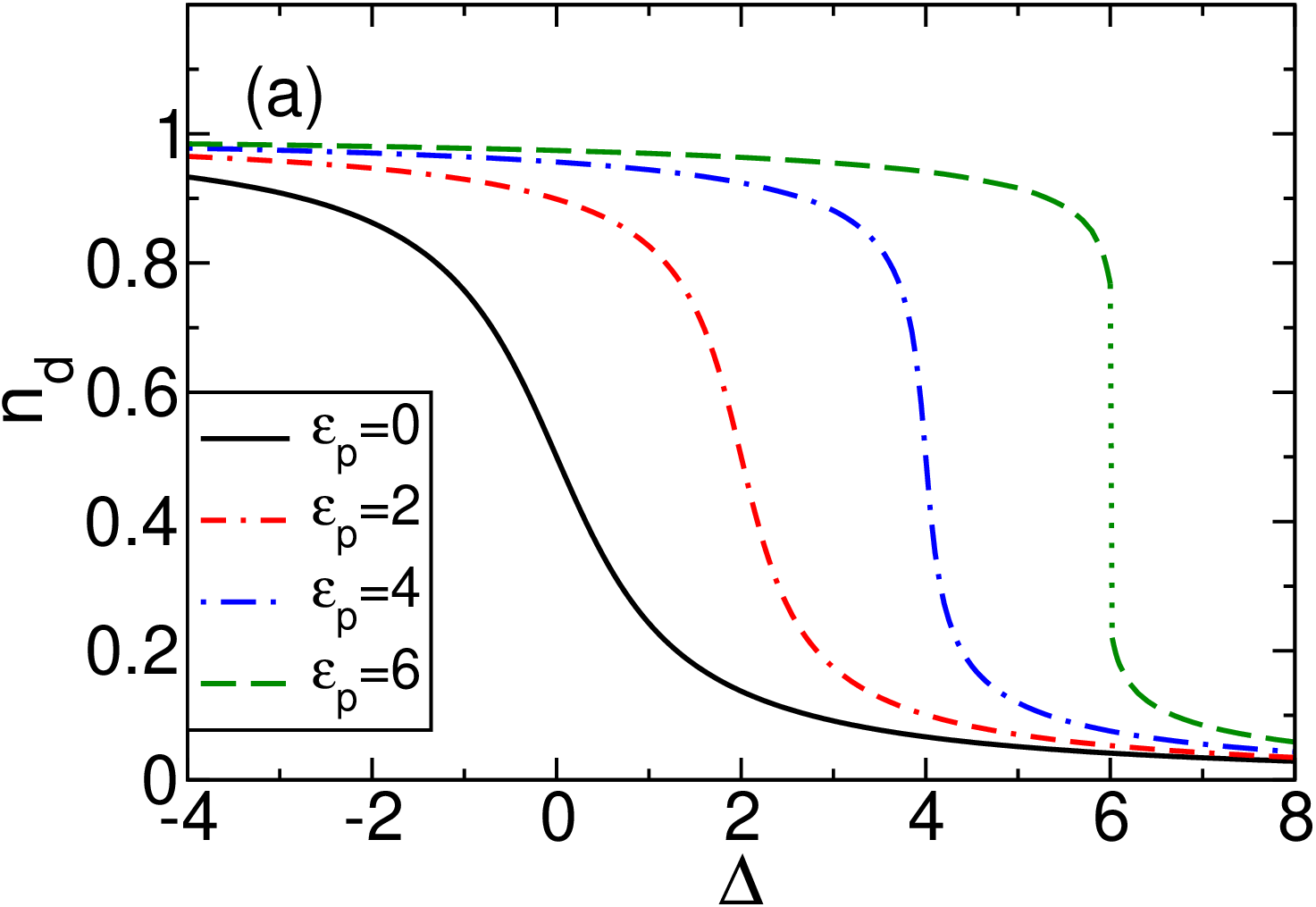}\hspace{0.1cm}
\includegraphics[width=0.32\linewidth]{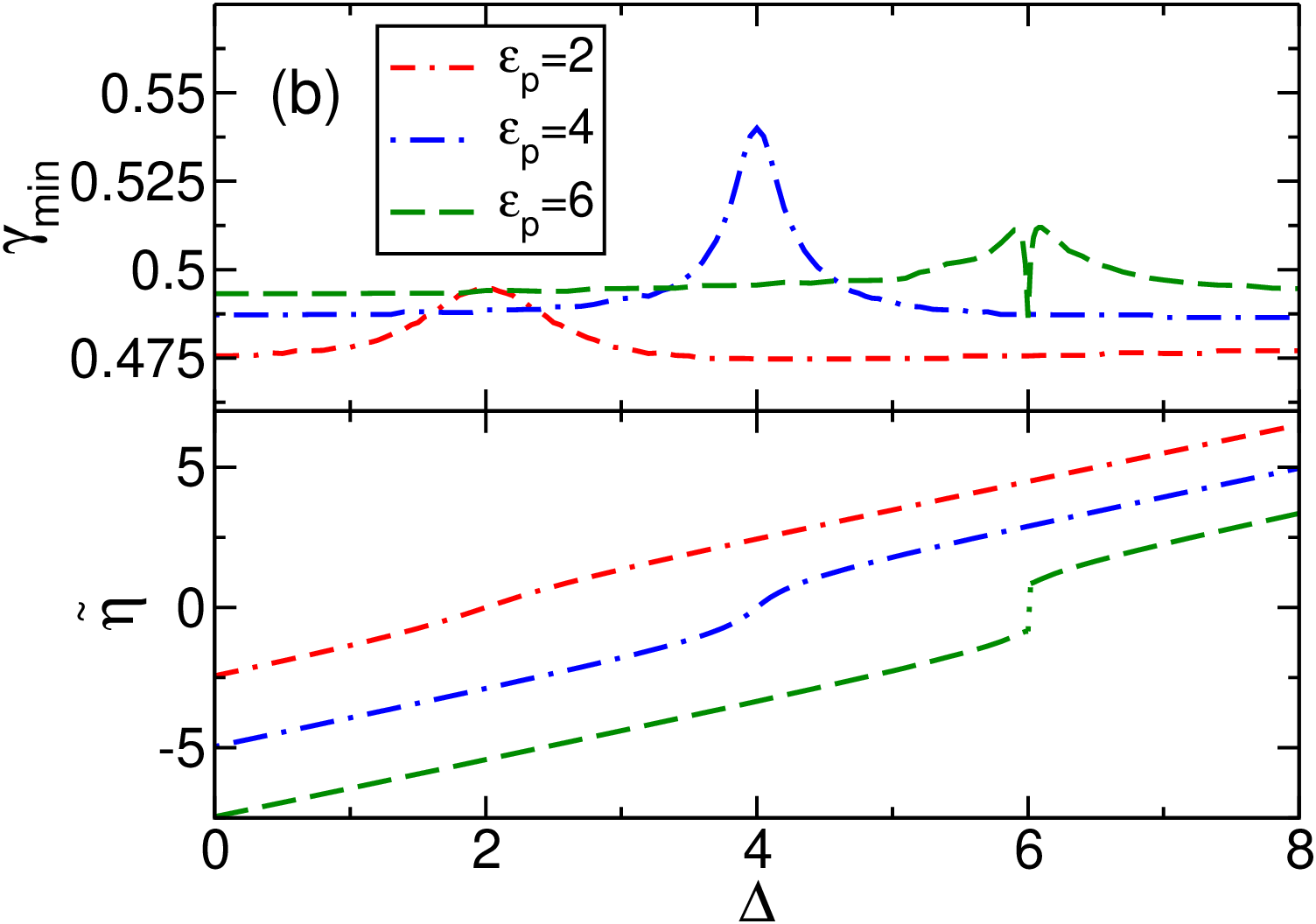}\hspace{0.1cm}
\includegraphics[width=0.32\linewidth]{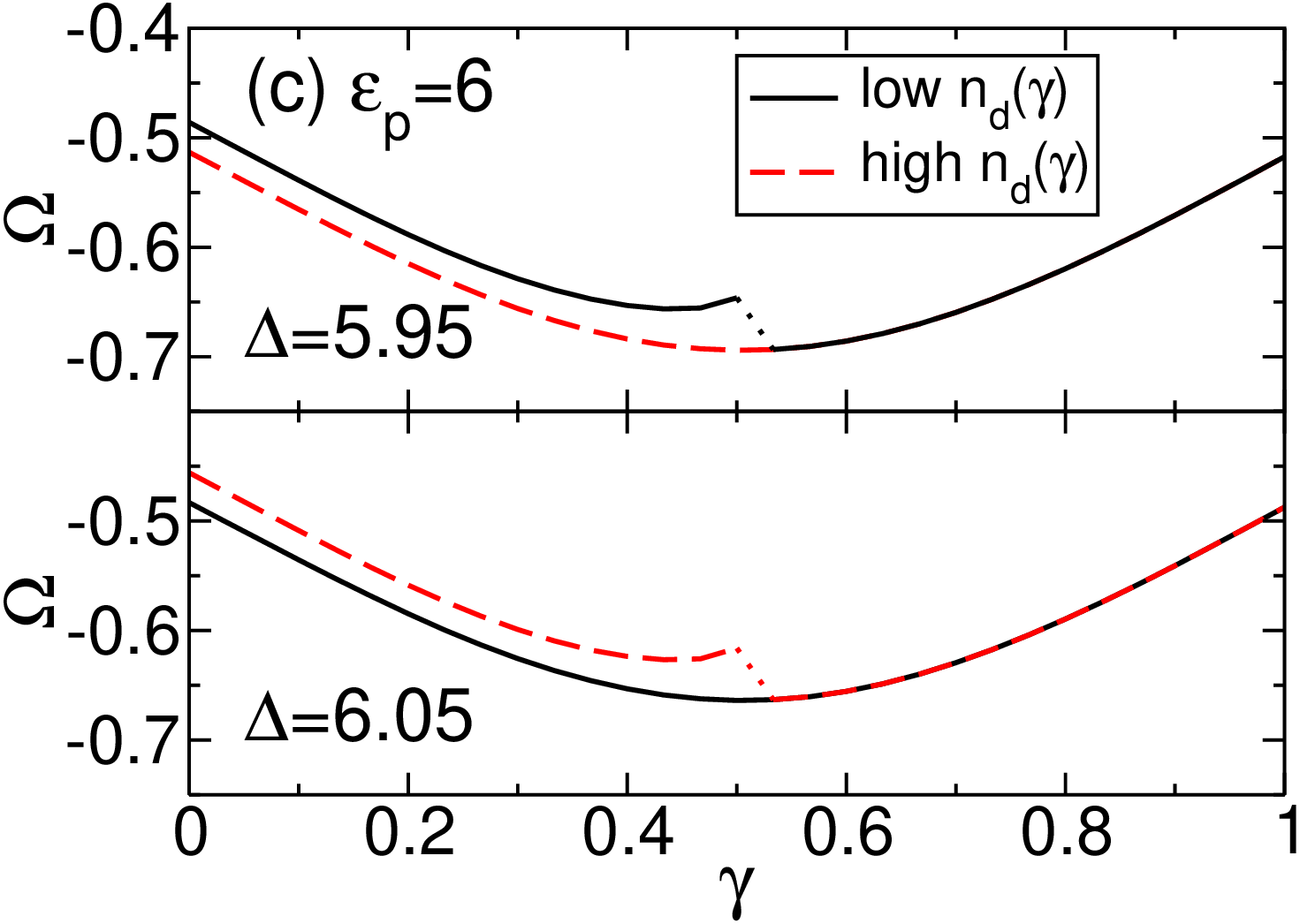}\\[0.2cm]
\includegraphics[width=0.33\linewidth]{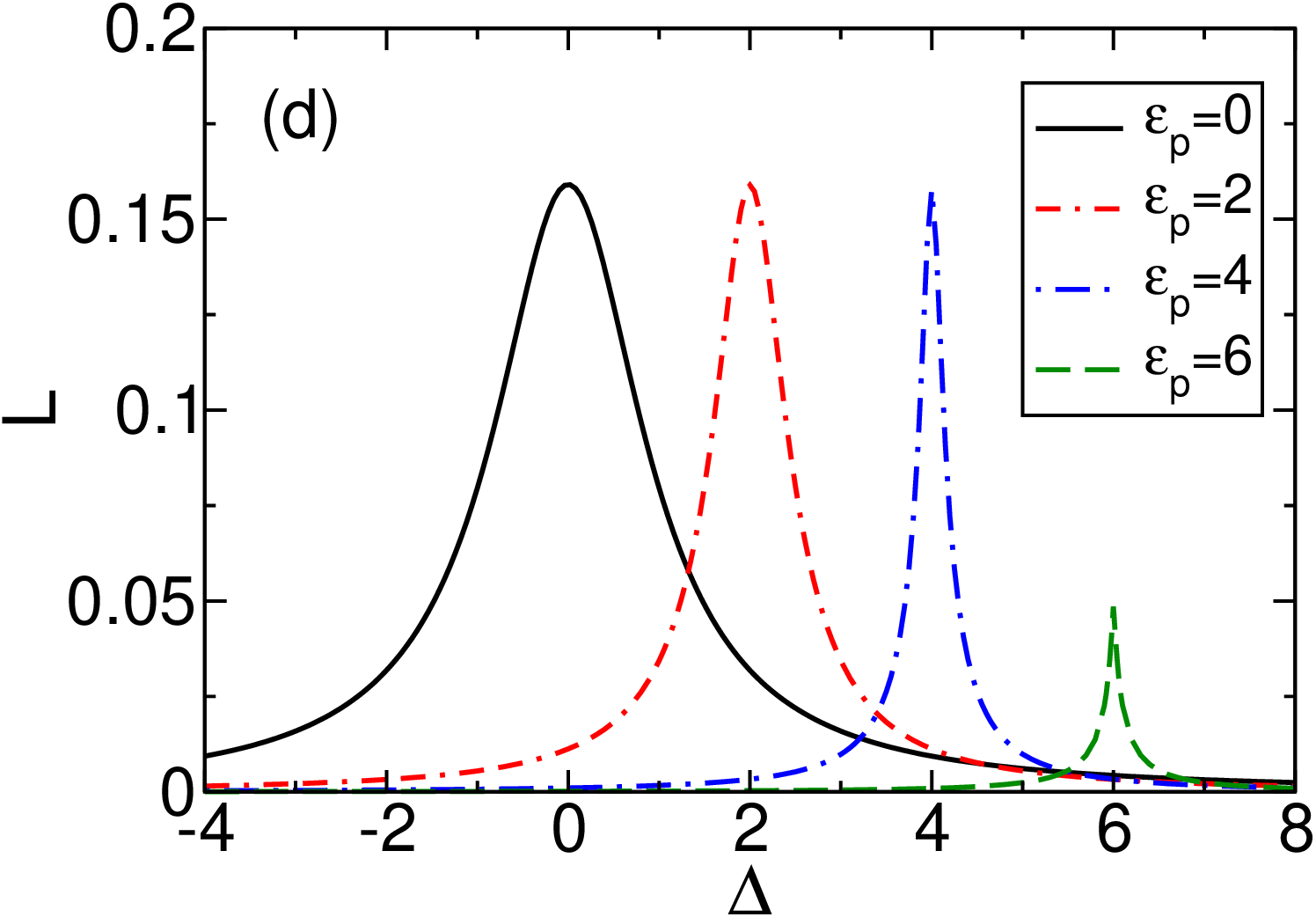}\hspace{0.1cm}
\includegraphics[width=0.32\linewidth]{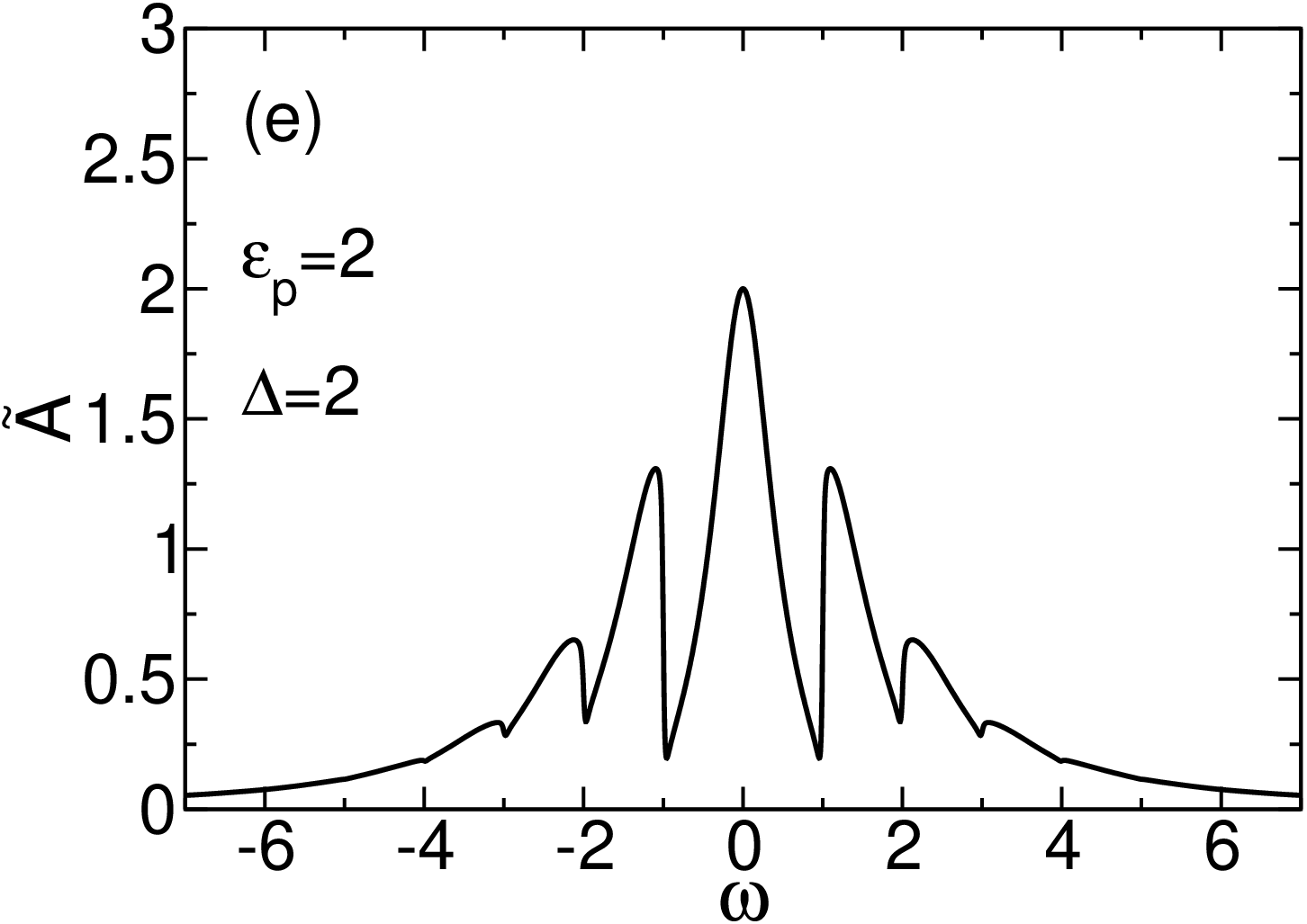}\hspace{0.1cm}
\includegraphics[width=0.32\linewidth]{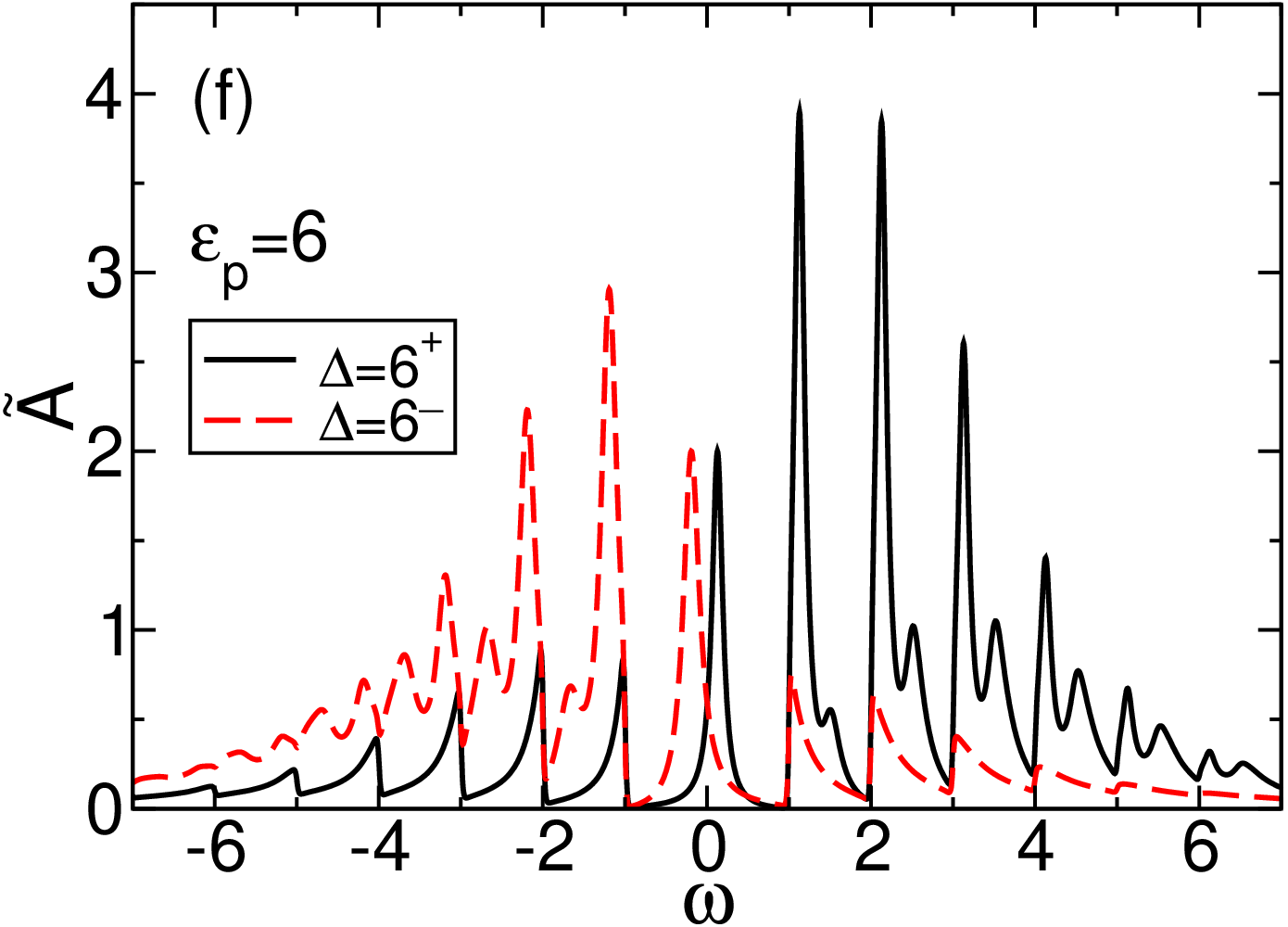}
\end{center}
\caption{For model parameters $\Gamma^{(0)}=1$, $T=0.01$, $\mu=0$ and $\Phi=0$. Panel (a): Dot occupation as a function of the bare dot level for several $\varepsilon_p$. Panel (b): Variationally determined $\gamma_{\mathrm{min}}$ and renormalized dot level as functions of the bare dot level $\Delta$. Panel (c): thermodynamic potential as a function of $\gamma$ for $\varepsilon_p=6$ and $\Delta$ in the vicinity of the discontinuous transition. Here we consider the lower (black solid line) or upper (dashed red line) root of the self-consistency equation for $n_d$. Panel (d): Linear conductance as a function of the bare dot level. Panel (e): Electronic spectral function for $\varepsilon_p=2$ at resonance. Panel (f): Electronic spectral functions for $\varepsilon_p=6$ and $\Delta$ slightly above ($\Delta=6^+$) and below ($\Delta=6^-$) the discontinuous transition.}
\label{fig2}
\end{figure*}

Again we compare the corresponding limit $\gamma=1$ to the result of the variational calculation while setting $\Delta=\varepsilon_p=1$ (see Fig.~\ref{fig1}(b)). In the former case, the dot level is renormalized by the polaron binding energy and represented by the zero-phonon peak at $\widetilde\Delta=\Delta-\varepsilon_p=0$. In addition we find pronounced peaks separated by $\omega_0$, signalling the emission of phonons by incident electrons and holes. The spectrum documents the formation of a long-living polaron state at the dot, with a mean number of phonons given by $g^2=1$.

For the same parameters, the variational calculation yields $\gamma_{\mathrm{min}}=0.81<1$ and we find a somewhat broader main peak and less spectral weight in the phonon sidebands ($\widetilde g^2=0.66$). Consequently, incoherent hopping transport through the dot takes place via an intermediate polaron state, whose spectral weight and lifetime are smaller than predicted by the complete ($\gamma=1$) Lang-Firsov calculation.

Figure~\ref{fig1}(c) finally shows the dot occupation and variational parameter as functions of the dot level $\Delta$ in the antiadiabatic case $\Gamma^{(0)}=0.1$, but for small EP coupling $\varepsilon_p=0.3$. In this regime, we find $\gamma_{\mathrm{min}}\approx0.7$. This is in good quantitative agreement with the result of La Magna and Deretzis,\cite{LD07} who applied a variational Lang-Firsov transformation to an effective electron model (cf. Fig. 2(b) in Ref. \onlinecite{LD07}). The above calculations show that, although the Lang-Firsov approach provides the correct physical mechanism, away from the very strong coupling limit, adiabatic corrections may not be neglected. 
%
%
%
\subsubsection{Intermediate dot-lead coupling regime}\label{SECintermediate}
\begin{figure*}[t]
\begin{center}
\includegraphics[width=0.32\linewidth]{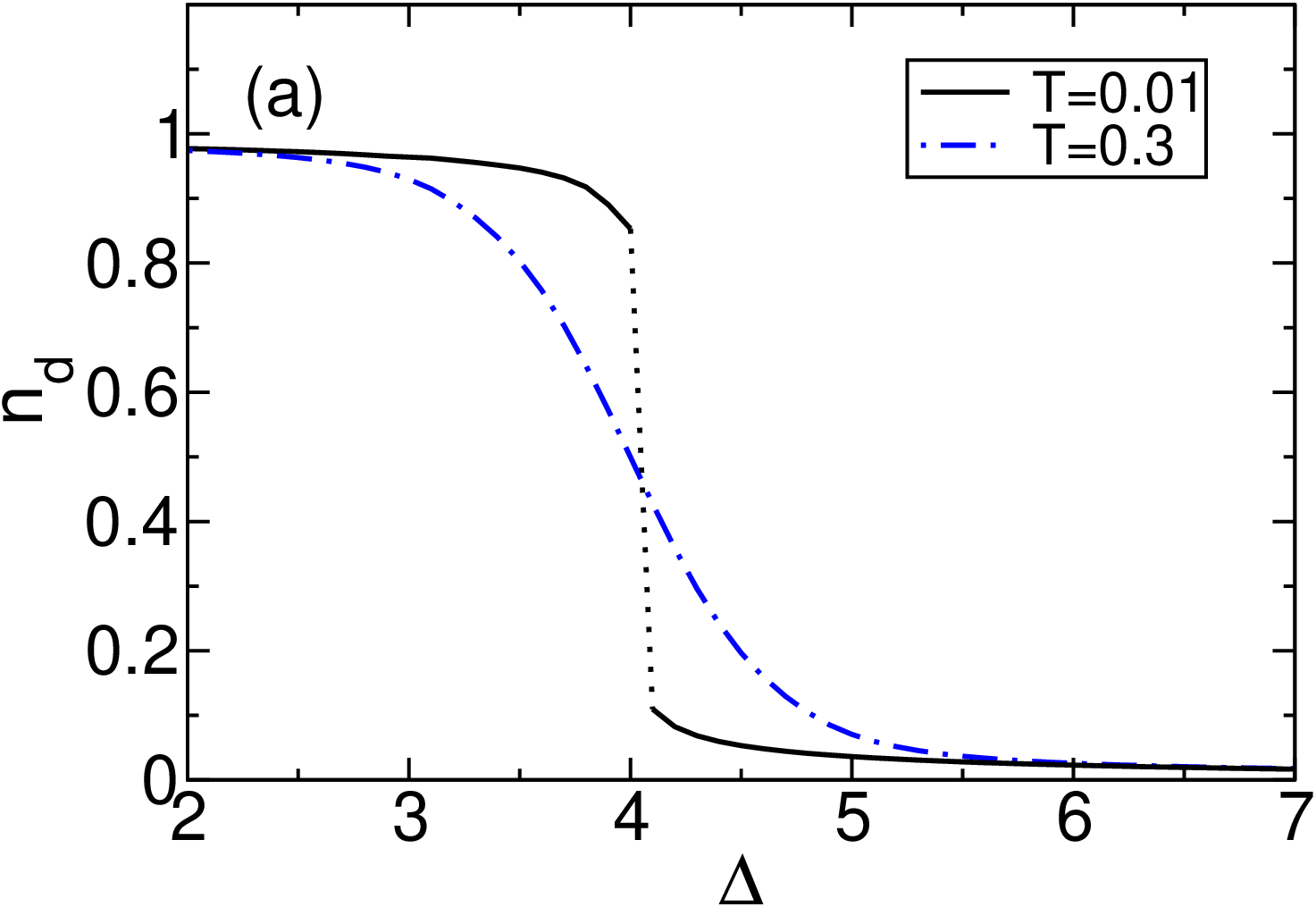}\hspace{0.1cm}
\includegraphics[width=0.32\linewidth]{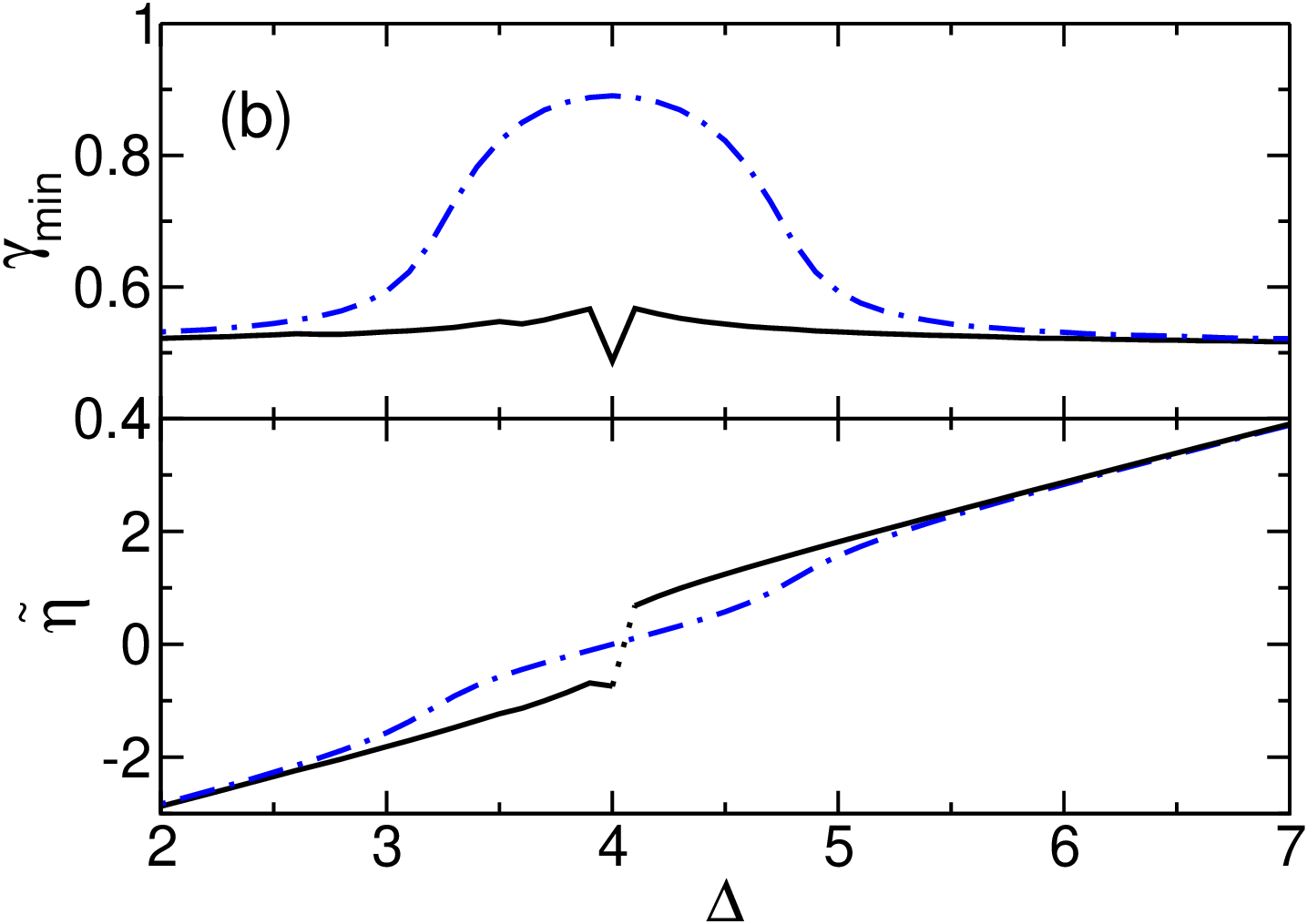}\hspace{0.1cm}
\includegraphics[width=0.32\linewidth]{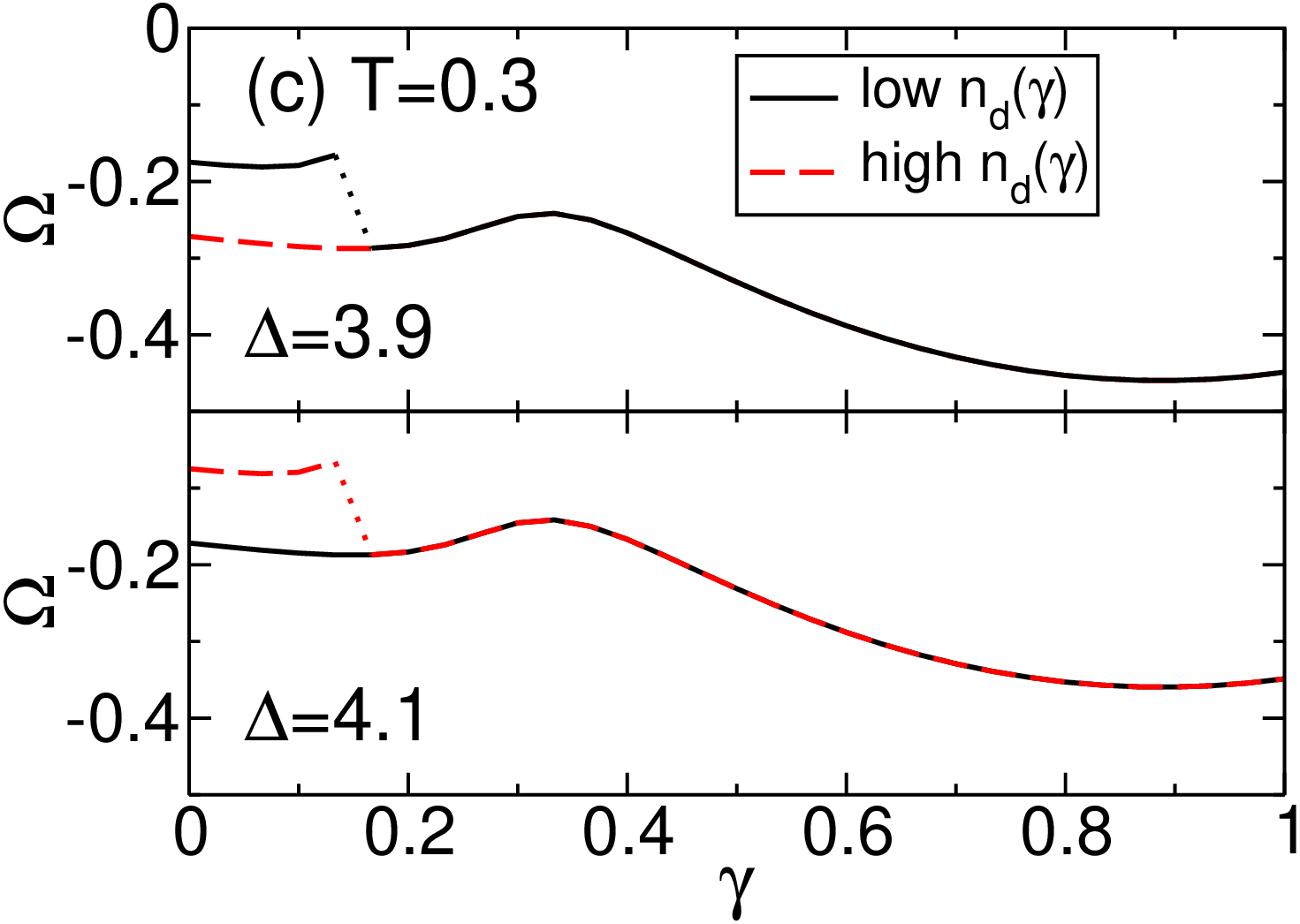}\\[0.2cm]
\includegraphics[width=0.32\linewidth]{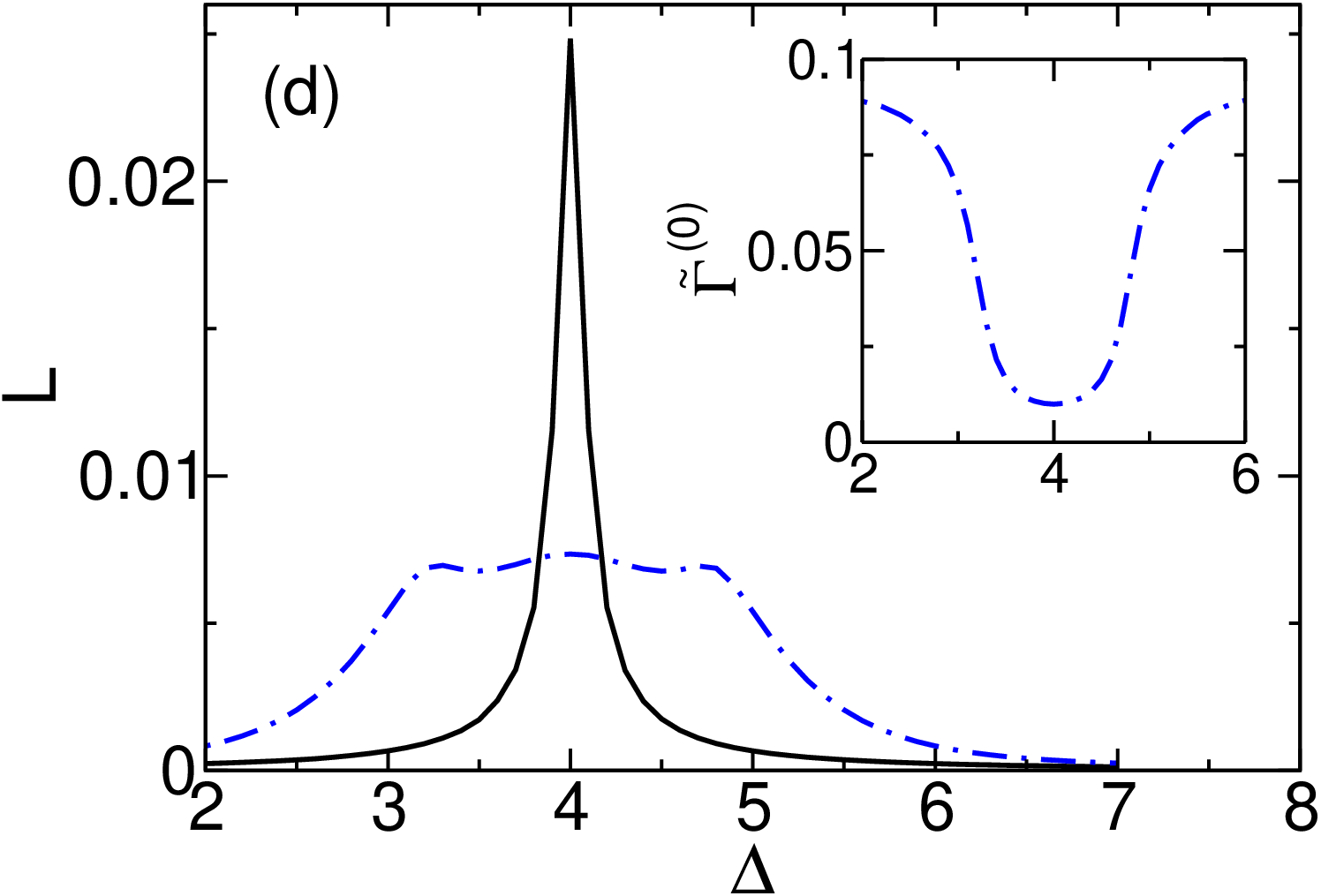}\hspace{0.1cm}
\includegraphics[width=0.32\linewidth]{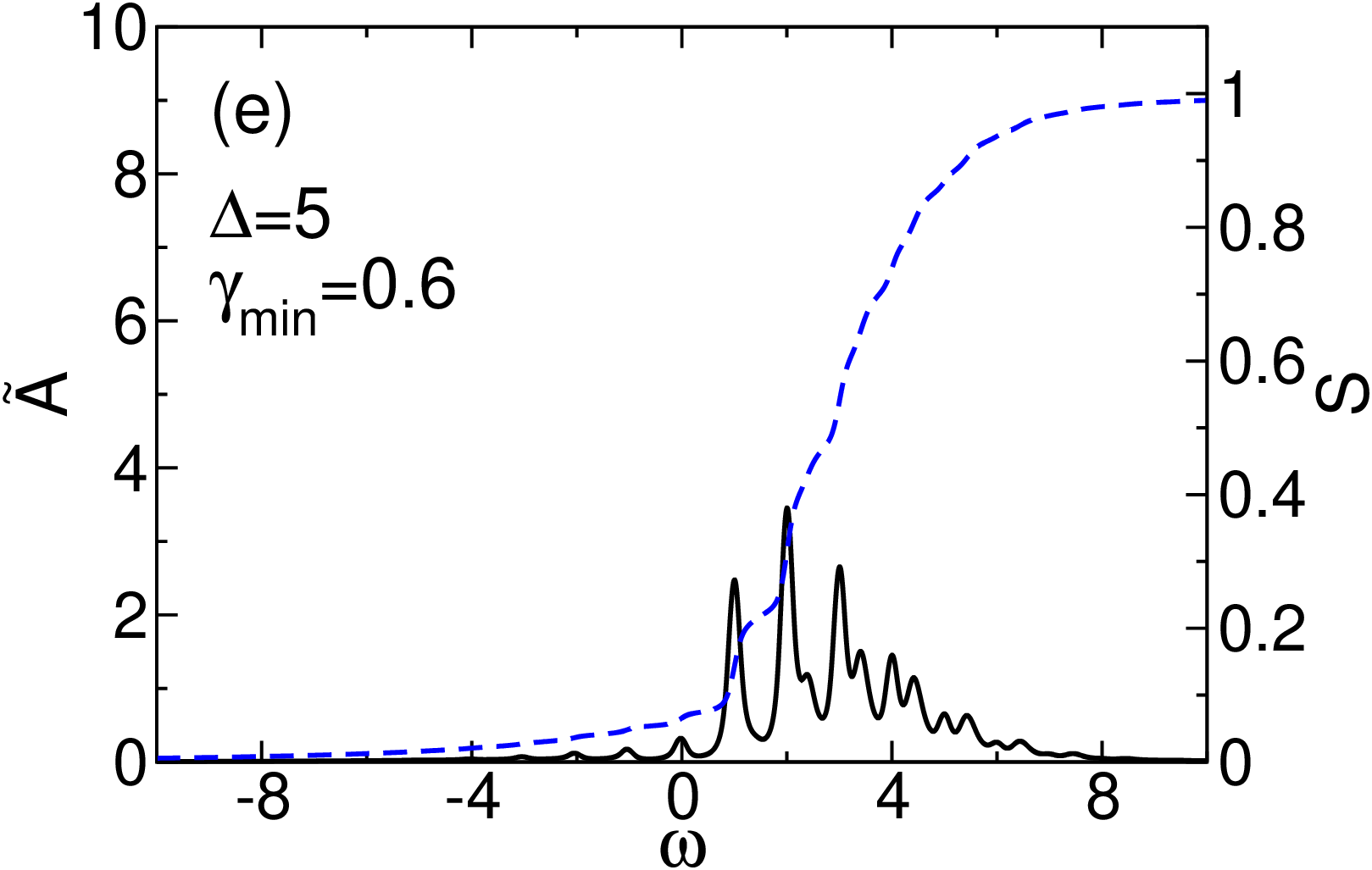}\hspace{0.1cm}
\includegraphics[width=0.32\linewidth]{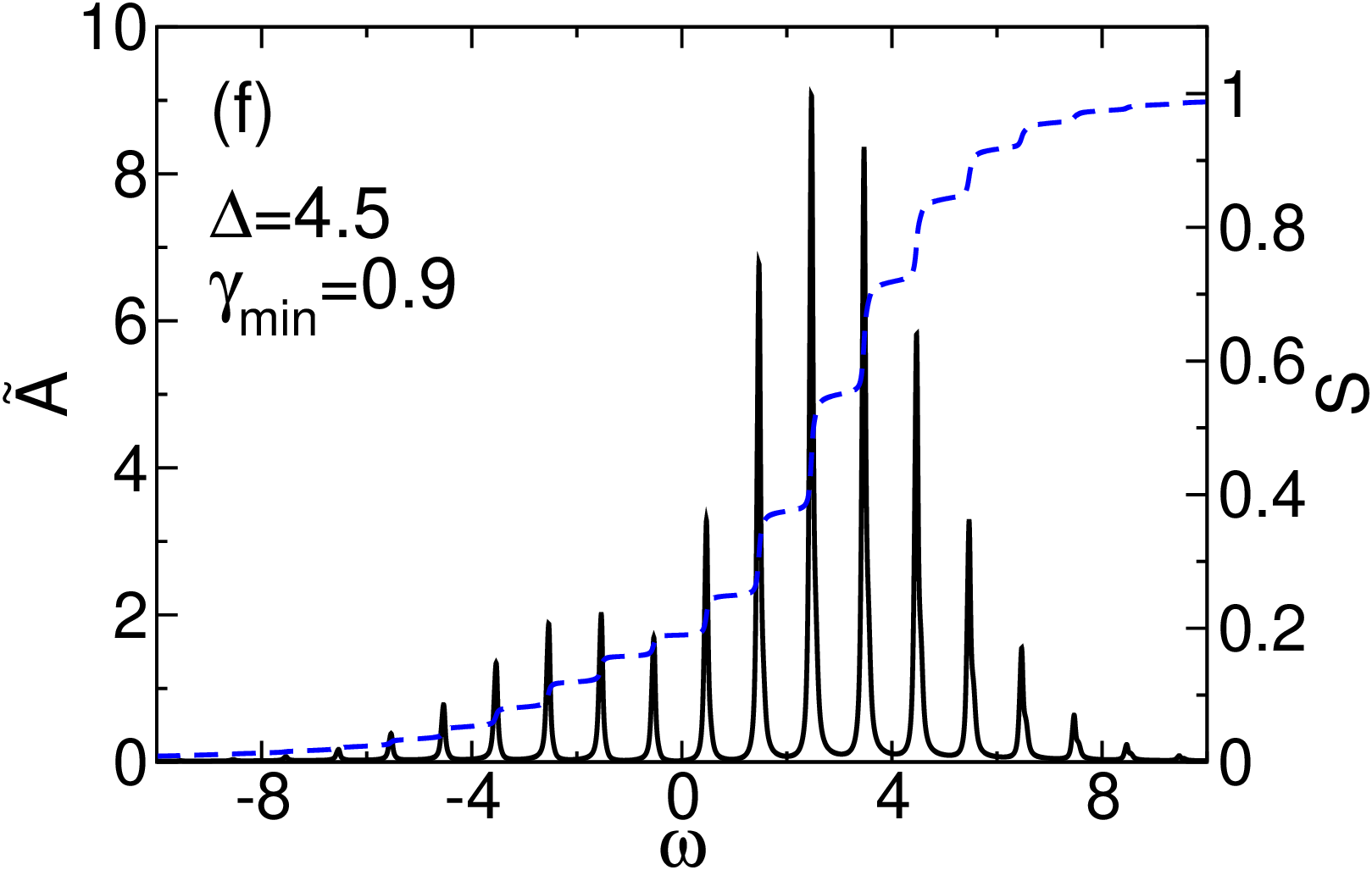}
\end{center}
\caption{For model parameters $\Gamma^{(0)}=0.3$, $\mu=0$, $\Phi=0$, $\varepsilon_p=4$ and several temperatures. Panel (a): Dot occupation as a function of the bare dot level. Panel (b): Variationally determined $\gamma_{\mathrm{min}}$ and renormalized dot level as functions of the bare dot level. Panel (c):  thermodynamic potential as a function of $\gamma$ for $T=0.3$ and $\Delta$ in the vicinity of the resonance. Here we consider the lower (black solid line) or upper (dashed red line) root of the self-consistency equation for $n_d$. Panel (d): Linear conductance as a function of the bare dot level. Inset: renormalized dot-lead coupling. Panels (e) and (f): Electronic spectral function $\widetilde A$ and integrated spectral weight $S$ for $T=0.3$ and $\Delta=5$ and $\Delta=4.5$, respectively.}
\label{fig3}
\end{figure*}
We now investigate the regime of comparable electronic and phononic time scales by setting $\Gamma^{(0)}=1$.
Figure~\ref{fig2} presents the results of the equilibrium calculation for zero to large EP coupling strengths. Shown here are, as functions of the bare dot level $\Delta$: the dot occupation $n_d$ (a), the variational parameter $\gamma_{\mathrm{min}}$ and the renormalized dot level $\widetilde\eta$ (b), the linear conductance $L$ (d). For fixed $\varepsilon_p$ and $\Delta$, Fig. \ref{fig2}(c) gives the thermodynamic potential as a function of $\gamma$ while Fig. \ref{fig2}(e) and Fig. (f) display the electronic spectral functions at $\Delta=\varepsilon_p$.

For $\varepsilon_p=0$, the self-energy (\ref{EQUSigma2}) is exact (black curves in Fig.~\ref{fig2}) and the rigid dot acts as a tunneling barrier. As $\Delta$ is lowered and the dot charges continuously, the linear conductance increases, reaching a maximum at $\Delta=0$, where the dot level aligns with the lead chemical potentials and resonant tunneling is possible. The width of the conductance resonance is determined by the electron lifetime $\Gamma^{(0)}$.

For finite $\varepsilon_p$, the variational parameter $\gamma_{\mathrm{min}}\approx 0.5$ and grows only slightly at $\Delta=\varepsilon_p$. As expected, for equal electronic and phononic time scales we are far from the weak coupling ($\gamma=0$) and strong coupling ($\gamma=1$) limits. As a consequence of the EP coupling, the charging transition from $n_d\approx0$ to $n_d\approx1$ shifts to higher $\Delta$ because of an overall lowering of the effective tunneling barrier. 
Due to the self-consistent mean-field coupling in Eq.~(\ref{EQUdefeta}), the transition becomes more rapid and even discontinuous for $\varepsilon_p>5$ (signalled by the dotted green lines). 
Here the system switches between two stable solutions of Eq.~(\ref{EQUnselfcons}) in analogy to the strong coupling results of Refs.~\onlinecite{GRN05,LD07}. 
Figure~\ref{fig2}(c) shows the thermodynamic potential as a function of $\gamma$ for $\varepsilon_p=6$ with $\Delta$ slightly below and above resonance. For $\gamma<0.55$, the effective mean-field coupling in Eq.~(\ref{EQUdefeta}) is so strong, that Eq.~(\ref{EQUnselfcons}) has two roots. For $\Delta<\varepsilon_p$, the global minimum of the thermodynamic potential, situated at $\gamma=0.5$, corresponds to high $n_d$. As $\Delta$ crosses the resonance, the roots change roles and the relevant $n_d$ jumps. An adiabatic phase transition from $n_d=0$ to $n_d=1$ was also found for a single electron at a vibrating quantum dot.\cite{AF08b,FWLB08} Rapid polaron formation and multistability are considered possible mechanisms for strongly nonlinear transport properties of molecular junctions such as NDC.\cite{LD07,GRN05,YGR07,*GNR08} 

From Fig.~\ref{fig2}(a) we see that, in case of a continuous transition, $n_d=0.5$ whenever $\Delta=\varepsilon_p$. As can be easily checked from Eq.~(\ref{EQUdefeta}), at this point the renormalized dot level resonates with the lead chemical potentials, i.e. $\widetilde\eta=0$ irrespective of $\gamma_{\mathrm{min}}$. Figure~\ref{fig2}(e) shows the corresponding electronic spectral function for moderate coupling $\varepsilon_p=\Delta=2$. Few ($\widetilde g^2=0.5$) broad sidebands signal phonon emission by either particles ($\omega>0$) or holes ($\omega<0$). The spectrum suggests that transmission remains coherent, but is governed by the slightly increased lifetime of the transient polaron state $\propto 1/\widetilde\Gamma^{(0)}$, with $\widetilde\Gamma^{(0)}=0.6$.
In case of a discontinuous charging, the dot level is shifted instantly across the resonance and there is no particle-hole symmetric situation, as is demonstrated by the spectral functions near the transition for $\varepsilon_p=6$ (see Fig.~\ref{fig2}(f)). Because $\widetilde g^2=1.5>1$, spectral weight is shifted from the narrow main peak to multiphonon states, reducing the tunneling rate in the off-resonant situation considerably (Franck-Condon blockade).

The effects of the EP coupling on the linear response of the quantum dot can be seen in Fig.~\ref{fig2}(d). Due to the rapid charging and the growing lifetime of the transient polaron the symmetrical conductance resonance shifts and narrows. This result coincides with the findings of Entin-Wohlmann {\it et al.} (Ref.~\onlinecite{EIA09}) and contradicts the $\varepsilon_p$-dependent broadening shown in the work of Mitra {\it et al.} (Ref.~\onlinecite{MAM04}). Note that in case of a continuous transition the maximum value of $L$ is independent of the EP coupling strength, because the dependence of $L$ on $\widetilde\Gamma^{(0)}$ cancels in the low temperature limit.\cite{EIA09,KO05} In the strong coupling limit, the resonance is skipped and the linear response signal lowers. 
In accordance with Refs.~\cite{EIA09,MAM04}, we find no side peaks in the linear conductance at low temperatures. This is due to ``floating" side bands \cite{MAM04} in the electronic spectral functions: for all $\Delta$ the phonon signatures are offset by $\omega_0$ below and above the lead Fermi level, as can be seen from Fig.~\ref{fig2}(f). Consequently they are not resolved in the low temperature linear response. This fact is missed by single particle approaches.\cite{LM02,*ZB03}  
%
%
%
\subsection{Equilibrium, high temperature}
In the following, we consider the effect of finite temperatures on the equilibrium properties of the quantum dot. We set $\Gamma^{(0)}=0.3$ and $\varepsilon_p=4$, thereby entering the strong coupling, nonadiabatic regime. 
Figure~\ref{fig3} shows the same quantities as Fig.~\ref{fig2}, but compares the low temperature result ($T=0.01$, black curves) to our findings for $T=0.3$, which, considering phonon energies in the order of $100$ meV,\cite{BHK81,*HPG97,Paea00} corresponds to room temperature.

Comparing the low temperature result in Fig.~\ref{fig3}(a) to the one for $\varepsilon_p=4$ in Fig.~\ref{fig2}(a) we see that the reduction of the bare electron tunneling rate increases the effective EP coupling strength in such a way that the charging transition becomes discontinuous. If we increase the temperature the transition becomes continuous again. 
As Fig.~\ref{fig3}(c) shows, for $T=0.3$ the optimal $\gamma$ is situated in a region where only a single root of Eq.~(\ref{EQUnselfcons}) exists (cf. Fig.~\ref{fig2}(c)).

Moreover, at high temperatures the Fermi edges of the leads soften. Thermally excited lead electrons see a considerably reduced injection gap so that the charging transition becomes wide spread. We know from Sec.~\ref{SEClimits} that in the strong coupling antiadiabatic regime at resonance, when phonon emission by electrons and holes is possible, the variational parameter $\gamma_{\mathrm{min}}$ comes close to unity. At finite temperatures $T\approx \omega_0$ absorption of free phonons by incident electrons opens additional inelastic transmission channels. Our ansatz accounts for this with $\gamma_{\mathrm{min}}$ approaching one at $\Delta\approx 4.5$ well above resonance. 
The polaron formation is signalled by two wiggles in the renormalized dot level. The impact on the linear conductance can be seen in Fig.~\ref{fig3}(d): in contrast to the low temperature result, we now find three peaks in $L$. 

Figures \ref{fig3}(e) and (f) compare the electronic spectral functions before and after the polaron formation. For $\Delta=5$ and $\gamma_{\mathrm{min}}\approx0.6$, nearly all spectral weight lies in a few overlapping emission signals situated above the chemical potential. Because at $T\approx\omega_0$ the floating condition mentioned in Sec.~\ref{SECintermediate} is relaxed, we find a small phonon peak at the chemical potential. That is why the conductance resonance broadens with respect to the low temperature result. For $\Delta\to4.5$, the phonon peaks are shifted away from the chemical potential. As $\gamma$ approaches one, the polaron life time $\propto1/\widetilde\Gamma^{(0)}$ is increased by one order of magnitude (see inset Fig.~\ref{fig3}(d)). Consequently, the peaks in the spectral function narrow and spectral weight is transfered to higher order phonon signals. The net linear response, being an average over transmission channels near the chemical potential, decreases and shapes the outer conductance peaks. At $\Delta=\varepsilon_p=4$ the narrow zero phonon peak crosses the usual resonance. We note that the maximum value of $L$ is smaller than in the low temperature calculation. 
%
%
%
%
%
%
%
%

\subsection{Nonequilibrium situation}
The most important experimental technique for the characterization of molecular junctions is IETS. Experiments can be subdivided into nonresonant and resonant tunneling scenarios (RIETS). In the former, the energy of the molecular ion (i.e. $\widetilde\eta$) lies far above the lead chemical potentials. Consequently, electron residence times are short and inelastic effects are small. In the latter, resonance is achieved via the application of a gate voltage and strong EP interaction is expected.  In both cases the current-voltage characteristics exhibit distinct features attributed to vibrational coupling at the junction. In analogy to the preceding sections, we will analyze the adiabatic and antiadiabatic limiting cases before considering equal phononic and electronic time scales.

\subsubsection{Limiting cases} \label{SEClimits2}
\begin{figure}[t]
\begin{center}
\hspace{-1.2cm}\includegraphics[width=0.45\textwidth]{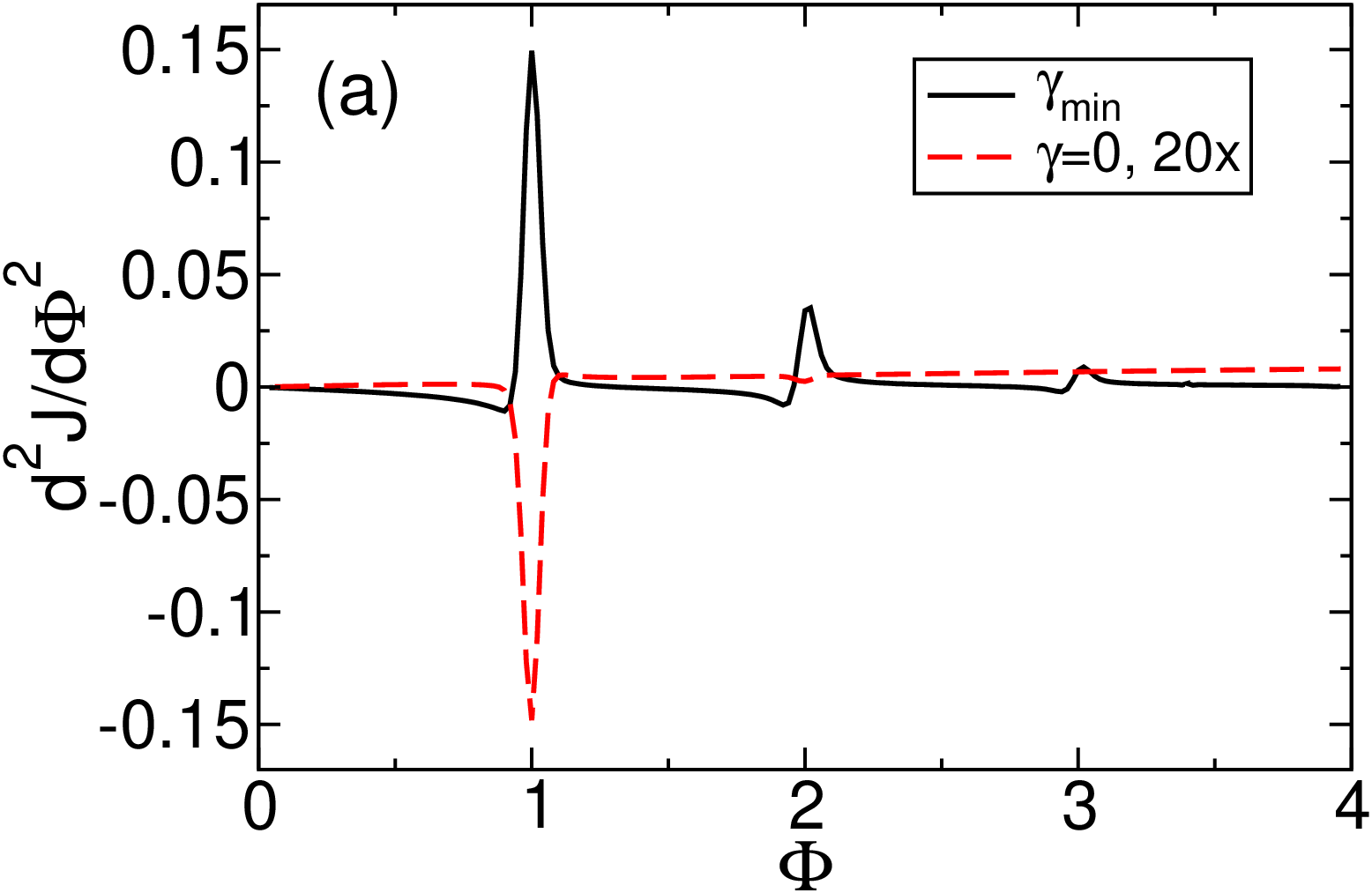}\\[0.2cm]
\includegraphics[width=0.48\textwidth]{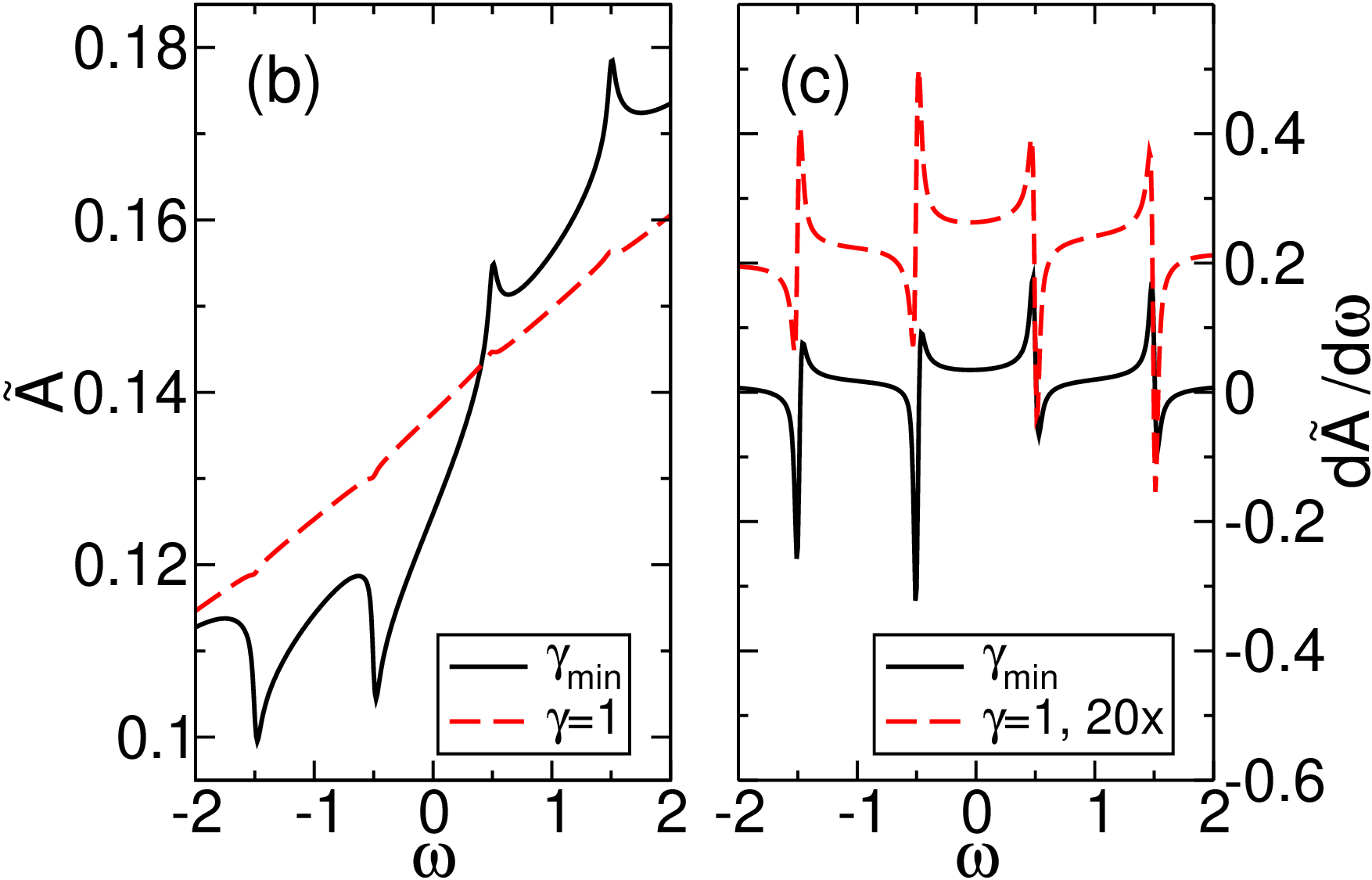}
\end{center}
\caption{For model parameters $T=0.01$, $\Gamma^{(0)}=10$, $\varepsilon_p=2$ and $\Delta=8$. Panel (a): Second derivative of the electron current as a function of the voltage bias for fixed $\gamma=0$ (scaled by a factor of $20$) and variationally determined parameter $\gamma_{\mathrm{min}}$, respectively. Panels (b) and (c): Electronic spectral functions and their first derivatives at $\Phi=\omega_0$.
}
\label{fig4}
\end{figure}
Figure~\ref{fig4}(a) shows the second derivative of the total electron current as a function of the voltage in the nonresonant ($\Delta=8$) adiabatic regime ($\Gamma^{(0)}=10$) for intermediate EP coupling strength ($\varepsilon_p=2$). For fixed $\gamma=0$ we find a single dip at $\Phi=\omega_0$, where $\widetilde\eta=6.8$. Here, phonon emission by incident electrons causes an additional inelastic tunneling current. 
Moreover, quasielastic processes involving the emission and subsequent absorption of a single phonon are no longer virtual, because the intermediate polaron state is only partially occupied.
The tunneling current (\ref{EQUcurrent}) is an integral over the energies of all incident and outgoing electrons and does not resolve the various tunneling processes. Therefore polaronic features are observed in the second derivative of $J$. 
As Persson showed,\cite{Pe88} the destructive interference of the elastic and quasielastic processes may overcompensate the positive inelastic contribution, leading to the dip in the IETS signal. In their SCBA analysis, Galperin {\it et al.} (Ref.~\onlinecite{GRN04}) demonstrated the strong qualitative dependence of this signature on the dot level $\Delta$ and the bare molecule-lead coupling $\Gamma^{(0)}$. Our ansatz allows for the polaronic renormalization of both these parameters: 
At $\Phi=\omega_0$ the variational calculation gives an optimal $\gamma_{\mathrm{min}}=0.3$ and the effective dot level is further lowered ($\widetilde\eta=6.4$ at $\Phi=\omega_0$). As can be seen from the electronic spectral function in Fig.~\ref{fig4}(b), the spectral weight of inelastic electron tunneling processes at $\omega\ge\Phi/2=0.5$ grows at the cost of the elastic transmission at $\omega=0$. As a consequence, the overall IETS signal now shows a pronounced peak at $\Phi=\omega_0$ (note the scaling of the curves in Fig.~\ref{fig4}(a)) and additional phonon features whenever the voltage crosses integer multiples of $\omega_0$. With the current being an integral over the quantum dot spectrum, the qualitative change in the one-phonon IETS signal can be traced back the first derivative of $\widetilde A(\omega)$,\cite{MTU03} which can be seen in Fig.~\ref{fig4}(c).
When going from $\gamma=1$ to $\gamma_{\mathrm{min}}=0.3$, the sum of the peak derivatives of $\widetilde A$ at $\omega=\mu_{L,R}=\pm\Phi/2$ changes sign, showing that the inelastic tunneling current outweighs the destructive interference of the elastic channels. 
\begin{figure}[p]
\begin{center}
	\hspace{-1.2cm}\includegraphics[width=0.8\linewidth]{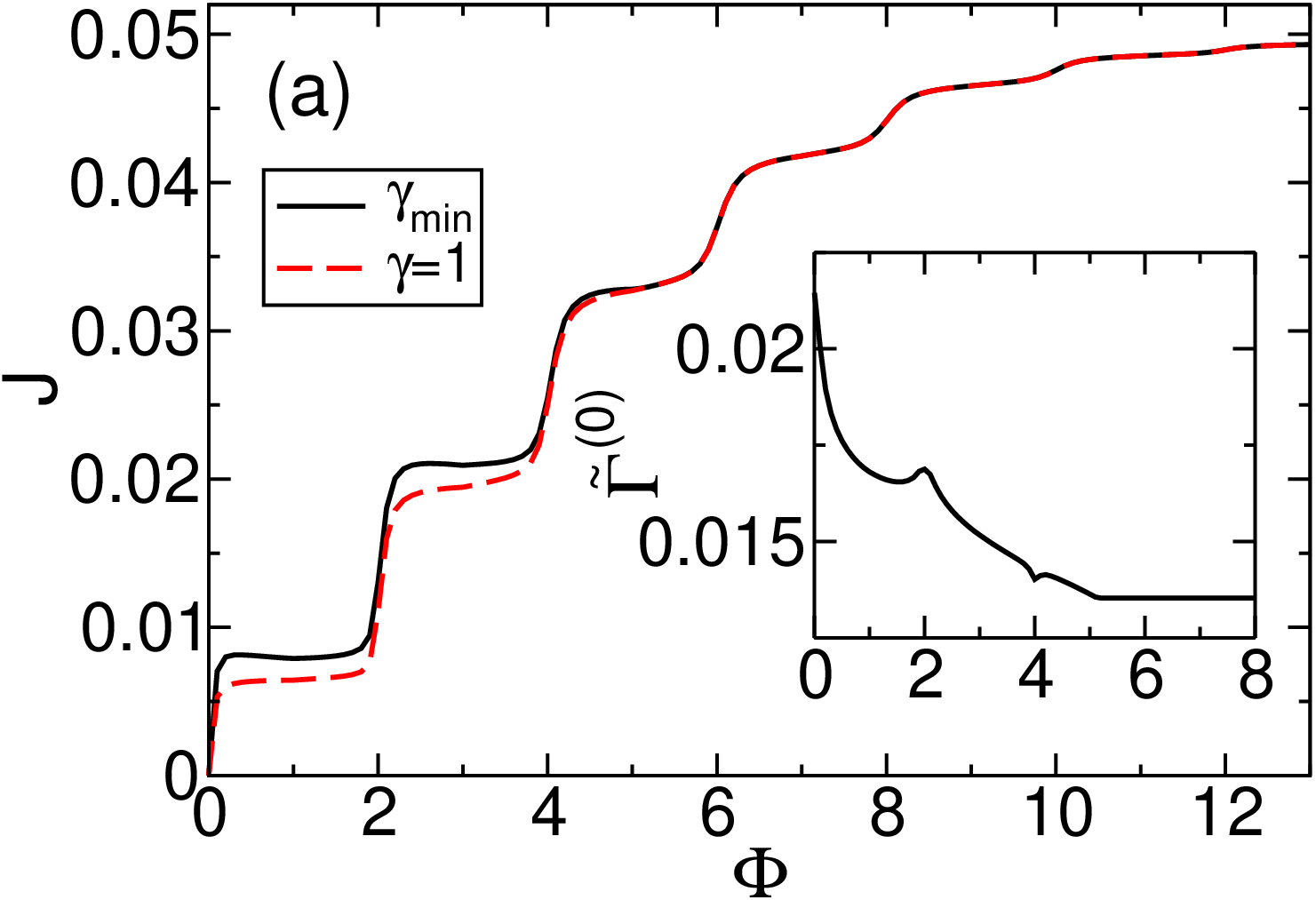}\\[0.2cm]
    	\hspace{-1.2cm}\includegraphics[width=0.8\linewidth]{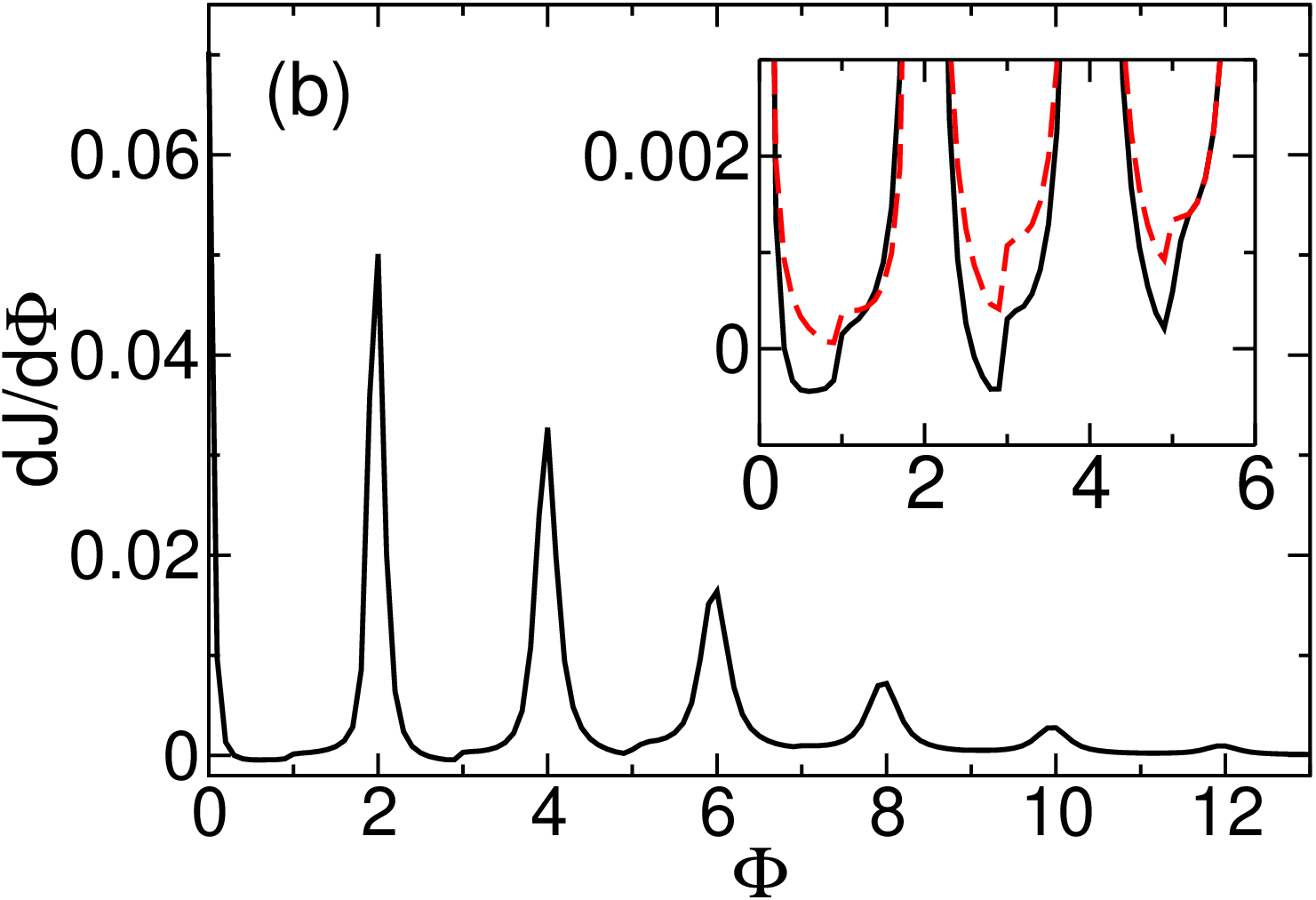}\\[0.2cm]
    	\includegraphics[width=0.8\linewidth]{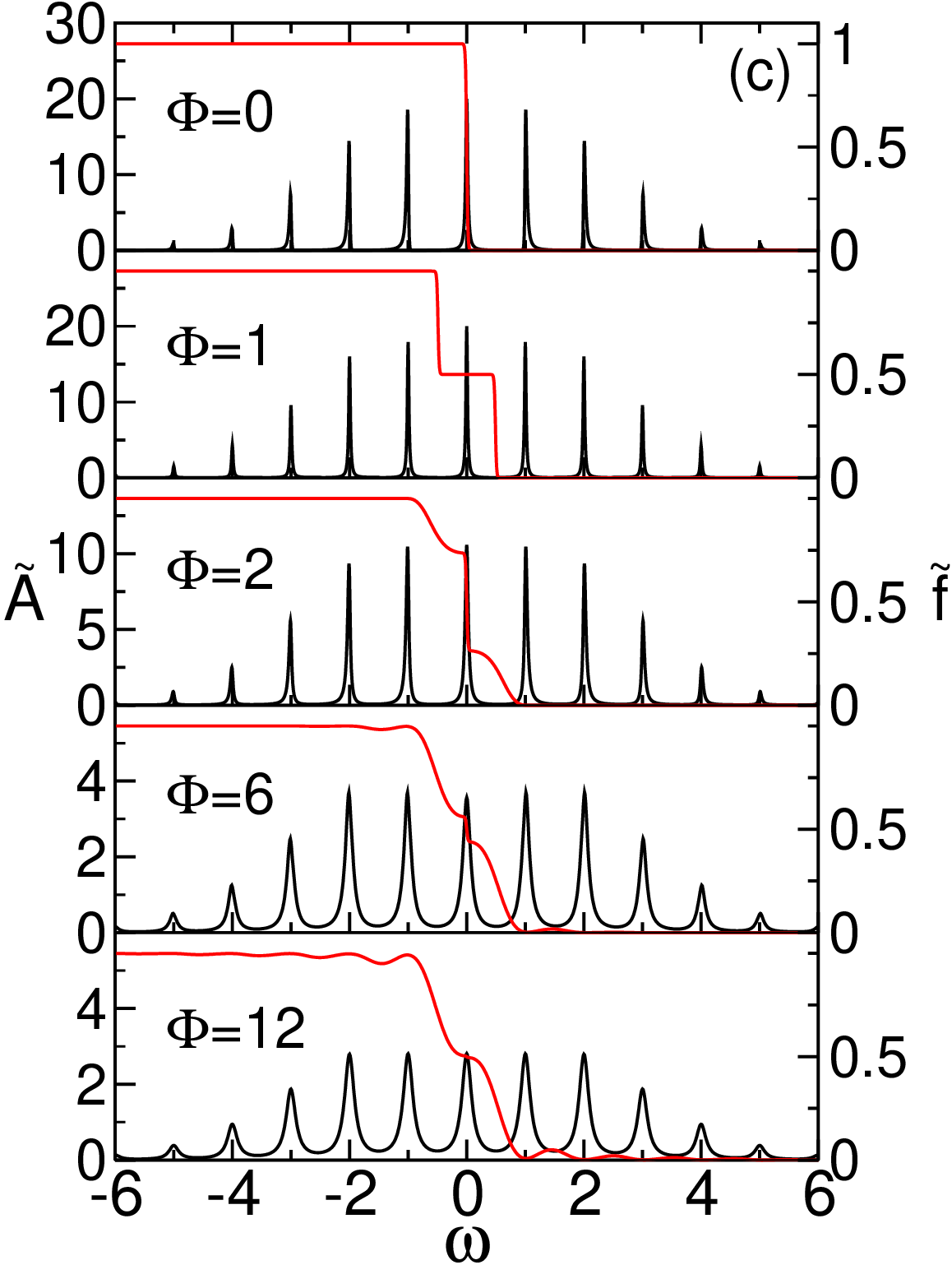}
\end{center}
\caption{For model parameters $T=0.01$, $\Gamma^{(0)}=0.1$, $\varepsilon_p=2$ and $\Delta=2$. Panel (a): Electron current as a function of the voltage bias, compared to the result with fixed $\gamma=1$. Inset: renormalized dot-lead coupling. Panel (b): Differential conductance as a function of the voltage bias. Inset: Zoom on the low-voltage region. Panel (c): Electronic spectral functions $\widetilde A$ and nonequilibrium electron distribution functions $\widetilde f$ for several voltages.
}
\label{fig5}
\end{figure}
\begin{figure}[p]
\begin{center}
	\hspace{-1.2cm}\includegraphics[width=0.8\linewidth]{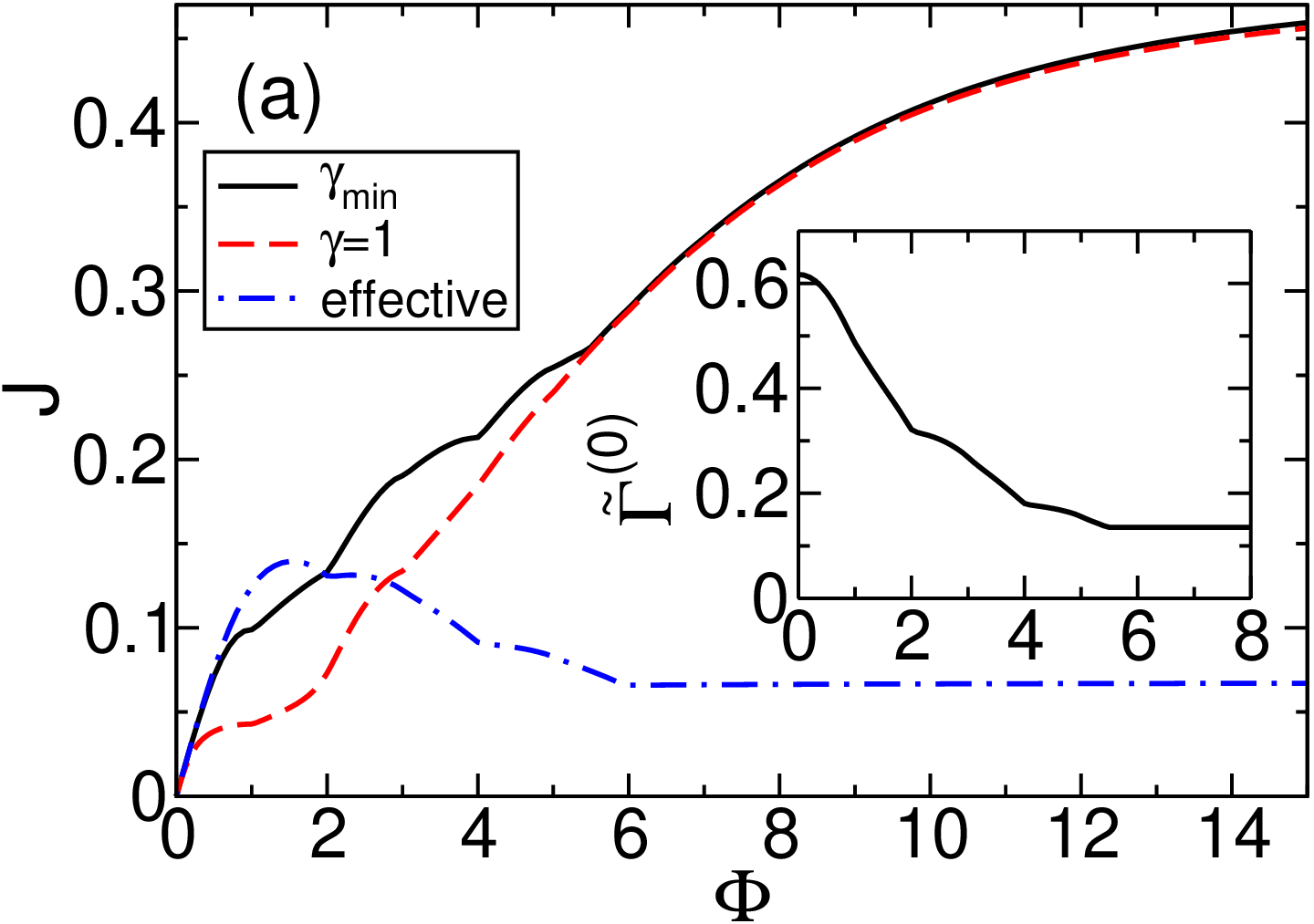}\\[0.2cm]
    	\hspace{-1.2cm}\includegraphics[width=0.8\linewidth]{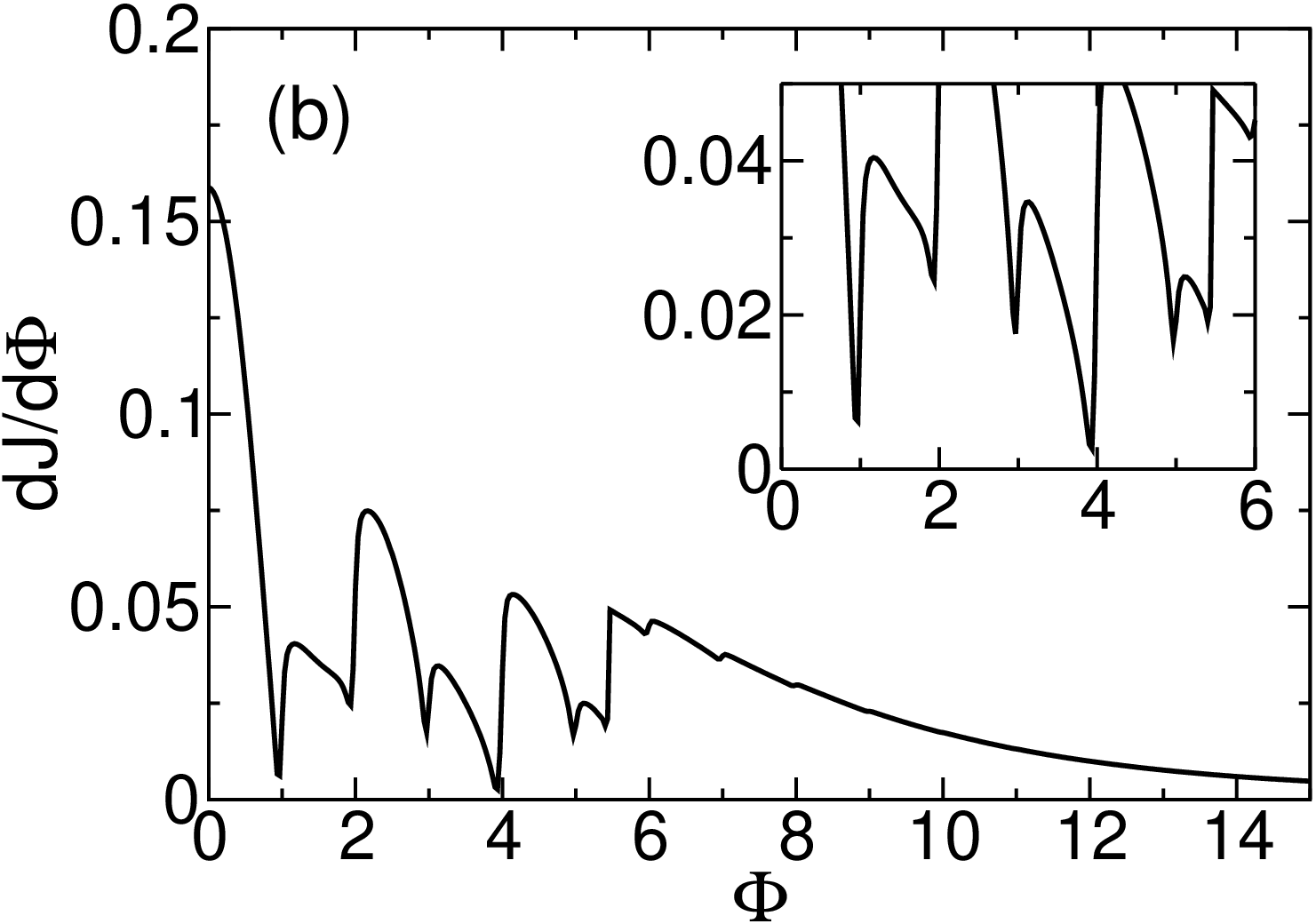}\\[0.2cm]
    	\includegraphics[width=0.8\linewidth]{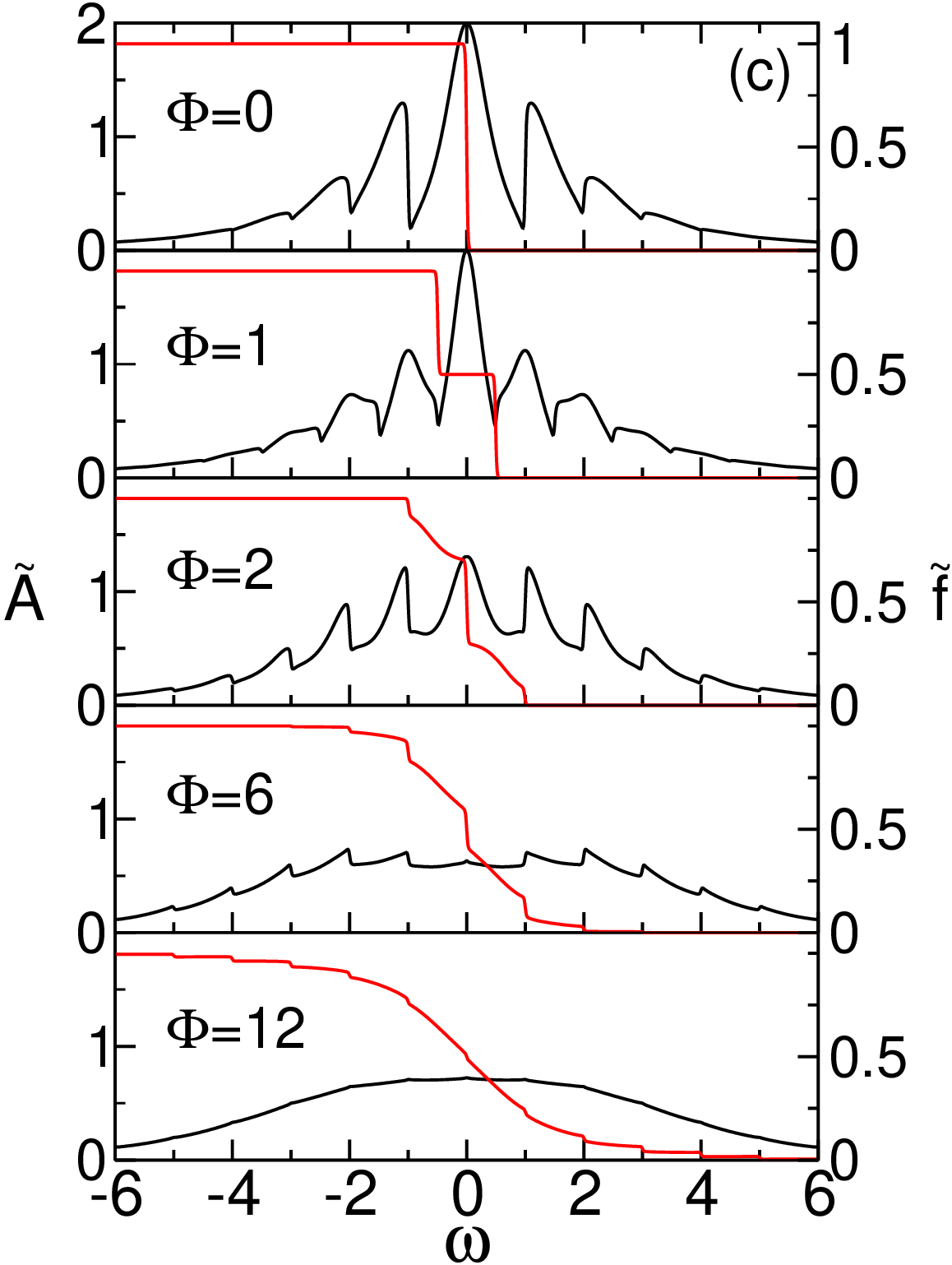}
\caption{$\Gamma^{(0)}=1$, $\Delta=2$, $\varepsilon_p=2$. Panel (a): Electron current as a function of the voltage for the variational calculation ($\gamma_{\mathrm{min}}$), compared to the result with fixed $\gamma=1$ as well as an effective electron model using renormalized parameters $\widetilde\Gamma^{(0)}$ and $\widetilde\eta$ determined by the variational calculation. Panel (b): Differential conductance as a function of the voltage bias. Inset: Zoom on the low-voltage region. Panel (c): Electronic spectral functions $\widetilde A$ and nonequilibrium electron distribution functions $\widetilde f$ for several voltages.}
\label{fig6}
\end{center}
\end{figure}

Figure~\ref{fig5}(a) and (b) present the total current and differential conductance as functions of the voltage in the resonant ($\Delta=2$) antiadiabatic regime ($\Gamma^{(0)}=0.1$) for intermediate EP coupling strength ($\varepsilon_p=2$). Because the voltage is raised symmetrically around the equilibrium chemical potential the dot occupation as well as the renormalized dot level $\widetilde\eta=0$ remain constant. 
Both, the variational calculation and the $\gamma=1$ case exhibit steps in the total current and pronounced peaks in the differential conductance whenever the voltage equals multiple integers of $2\omega_0$. Here resonant tunneling through phononic sidebands becomes possible. At $\Phi\approx 12$, the current saturates because now the so-called "Fermi window" $\omega\in[-\Phi/2,+\Phi/2]$ encompasses all phonon side bands (see. Fig. 5(c)). In the low-voltage region $\Phi<4$, the optimal variational parameter differs considerably from one ($\gamma_{\mathrm{min}}\approx0.9$), thereby increasing the overall weight of the relevant few-phonon inelastic tunneling channels.
As a consequence, the low-voltage current is larger than in the $\gamma=1$ case. Nevertheless, the growth of $\gamma_{\mathrm{min}}$ along a current plateau dynamically shifts spectral weight from the corresponding resonant inelastic channel to higher lying bands outside the Fermi window. As can be seen from the inset of Fig.~\ref{fig5}(b), the differential conductance is negative, which is in accordance with the polaron induced NDC found by La Magna and Deretzis.\cite{LD07} Only when an upward step (peak in $\mathrm{d}^2J/\mathrm{d}\Phi^2$) signals the opening of a nonresonant inelastic channel, the differential conductance becomes positive again.
\subsubsection{Intermediate dot-lead coupling regime} 
We now turn to the regime of equal electronic and phononic time scales, setting $\Gamma^{(0)}=1$ and keeping $T=0.01$ and $\varepsilon_p=2$ fixed.
First, we hold $\Delta=2$ at resonance, starting with $\gamma_{\mathrm{min}}=0.5$ and $n_d=0.5$ in equilibrium (cf. Fig.~\ref{fig2}). 
Figure~\ref{fig6}(a) presents the corresponding current-voltage characteristics. We compare the result of the variational calculation (black solid lines) to the case with fixed $\gamma=1$ (blue dashed lines) and to an effective electron model (red dash-dotted lines). 
The latter is obtained by setting $g=0$ in Eqs.~(\ref{EQUSigmalessfourier}) and (\ref{EQUGammaresult}) and inserting for $\Gamma^{(0)}_a$ the renormalized dot-lead coupling $\widetilde \Gamma^{(0)}$ resulting from the variational calculation. It is comparable to earlier works where the averaging over the phonon state leads to an effective electron Hamiltonian.\cite{LD07,LDV09}

With growing voltage the variational parameter steadily increases and approaches one in the high-voltage limit $\Phi>6$. The elastic transmission rate $\widetilde \Gamma^{(0)}$ shown in the inset of Fig.~\ref{fig6}(a) decreases accordingly. It exhibits steps at integer multiples of $2\omega_0$, suggesting that the polaron formation is especially rapid whenever a new resonant inelastic channel is accessible. The electronic spectral functions in Fig.~\ref{fig6}(c) show that spectral weight is shifted from the zero-phonon peak to the overlapping phonon side bands.
\begin{figure}[p]
\begin{center}
	\hspace{-0.7cm}\includegraphics[width=0.775\linewidth]{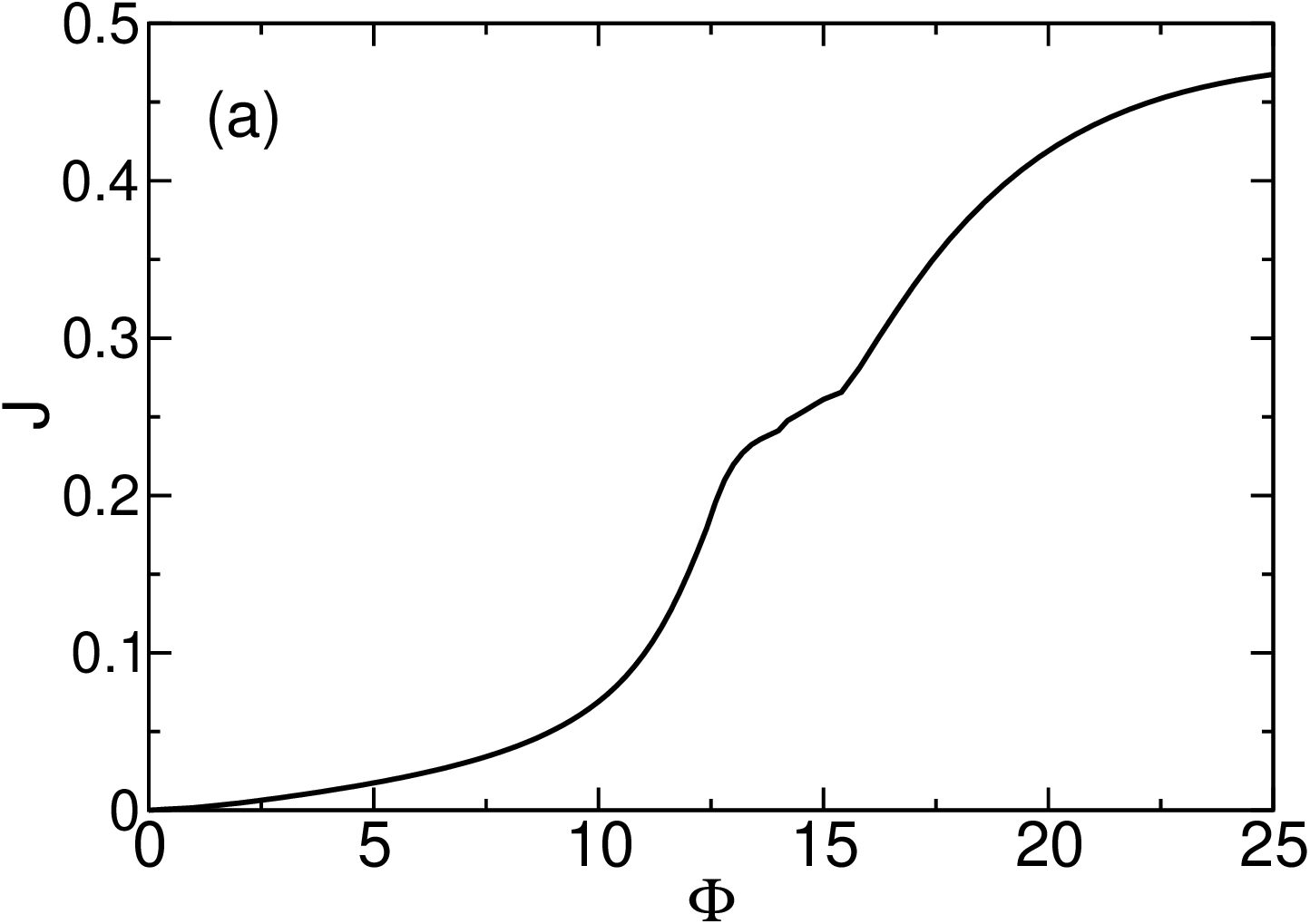}\\[0.2cm]
    	\hspace{-0.7cm}\includegraphics[width=0.775\linewidth]{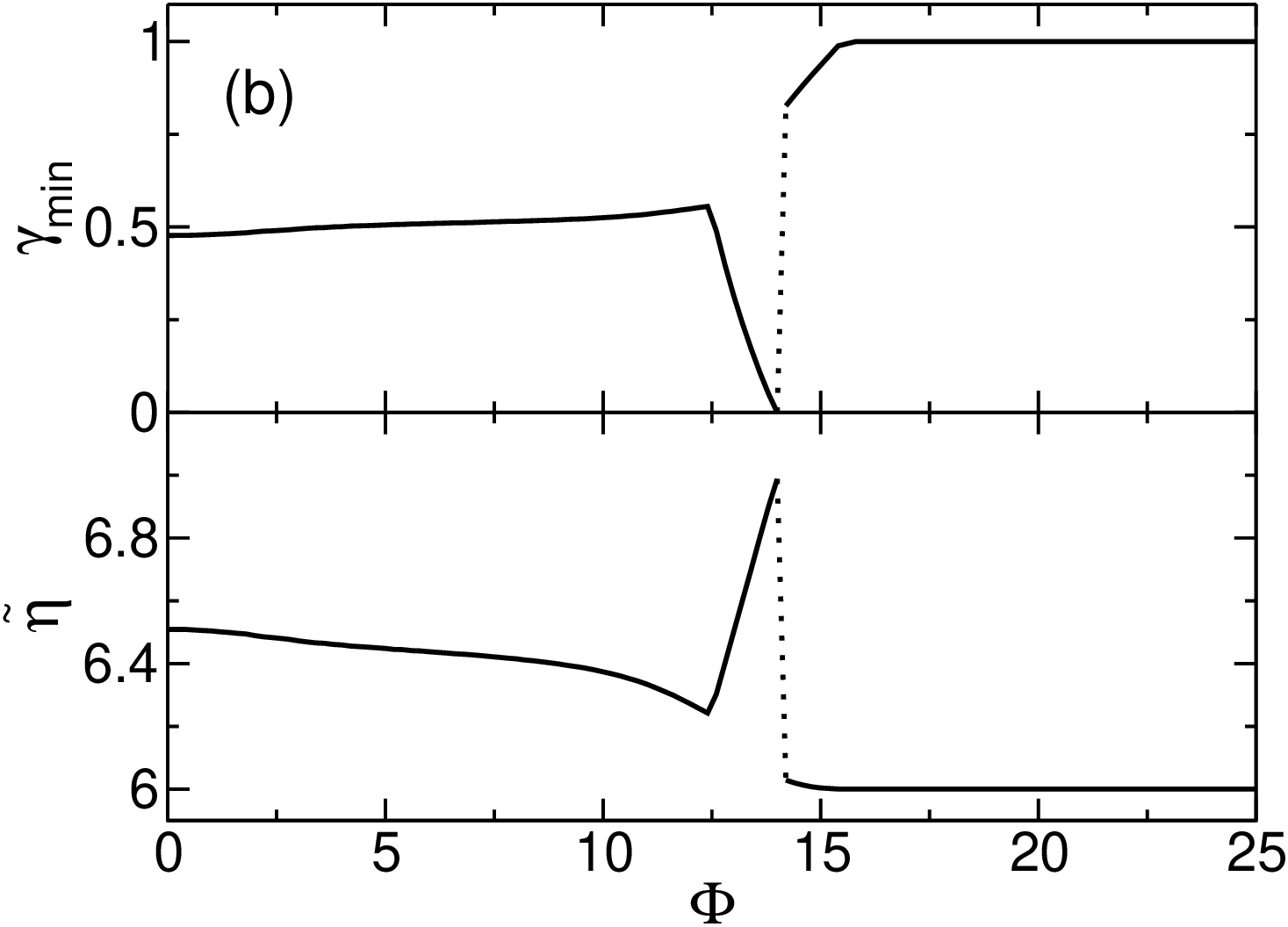}\\[0.2cm]
    	\includegraphics[width=0.85\linewidth]{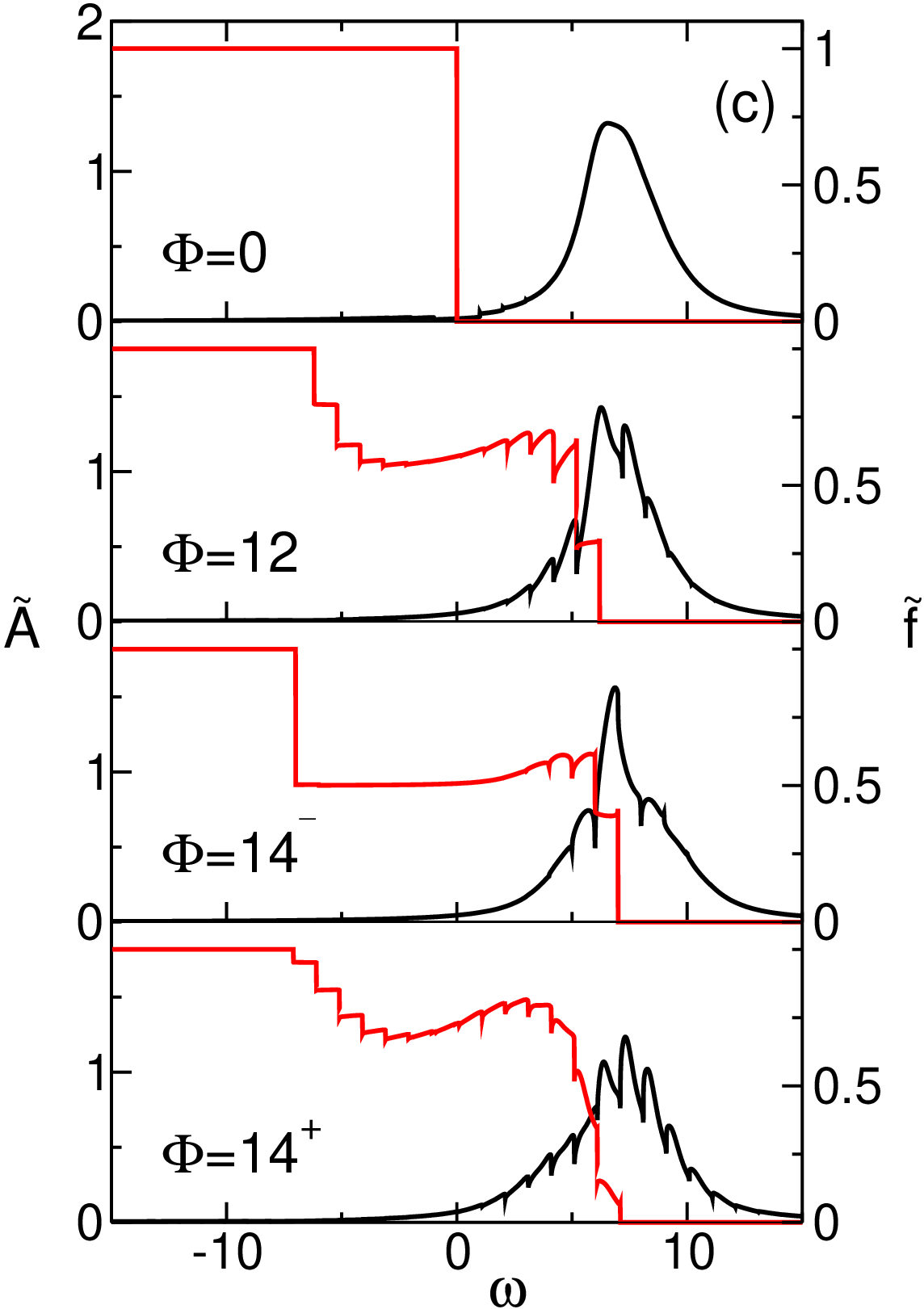}
\caption{$\Gamma^{(0)}=1$, $\Delta=8$, $\varepsilon_p=2$. Panel (a): Electron current as a function of the voltage. Panel (b): Variationally determined parameter $\gamma_{\mathrm{min}}$ and renormalized dot-level as functions of the voltage. Panel (c): Electronic spectral functions $\widetilde A$ and nonequilibrium electron distribution functions $\widetilde f$ for several voltages, e.g. slightly below ($\Phi=14^-$) and above ($\Phi=14^+$) the jump in $\gamma_{\mathrm{min}}$.}
\label{fig7}
\end{center}
\end{figure}

The current-voltage characteristics of the interacting results ($\gamma_{\mathrm{min}}$ and $\gamma=1$) contain signatures of both limiting cases discussed in Sec.~\ref{SEClimits2}, as can be seen from the differential conductance in Fig.~\ref{fig6}(b). As before, at voltages corresponding to integer multiples of $2\omega_0$, steps in the current (peaks in the conductance) signal the onset of resonant inelastic tunneling. These steps are considerably broadened and overlap with the onset of nonresonant inelastic tunneling. As a consequence, the polaron induced renormalization of the resonant channel is compensated and, in contrast to the low-voltage antiadiabatic regime, $\mathrm{d}J/\mathrm{d}\Phi$ remains strictly positive. 

The effective electron model overestimates the current in the region $\omega_0<\Phi<2\omega_0$. Since the spectrum contains no phonon side bands, for $\Phi>2\omega_0$ the decrease of the elastic tunneling rate $\propto \widetilde \Gamma^{(0)}$ is not compensated by resonant or nonresonant inelastic transmission processes. Consequently, we find a considerably lower maximum current and, in accordance with the results of La Magna and Deretzis,\cite{LD07} NDC in the intermediate-to-high voltage region. We conclude that the polaron induced renormalization of the dot-lead coupling is indeed a possible mechanism for NDC. Yet, the effective electron calculation misses the spectral features that are essential for electron transport at voltages exceeding $\omega_0$. The interplay of several inelastic transmission channels may heavily reduce or, for $\Gamma^{(0)}\gtrsim \omega_0$, even prevent the occurrence of NDC.
Another interesting consequence of the dynamic polaron formation can be observed in the high voltage regime, where a crossover from nonresonant to resonant transport takes place. We keep the above system parameters, but start from the nonresonant equilibrium situation with $\Delta=8$. The result is presented in Fig.~\ref{fig7}. As the voltage is raised, the variational parameter as well as the effective dot level remain nearly constant and transport takes place via nonresonant inelastic tunneling. At $\Phi=12.4$ the chemical potential of one lead resonates with $\widetilde\eta=6.2$, causing a broad step in the total current. When the voltage is raised further, the system maximizes its kinetic energy by decreasing the polaronic shift in such a way, that $\widetilde\eta$ stays locked to the lead chemical potential (see Fig.~\ref{fig7}(b)). As the spectral functions in Fig.~\ref{fig7}(c) suggest, this happens at the cost of the inelastic transmission channels. As soon as $\gamma_{\mathrm{min}}=0$ and resonance of the zero-phonon level can no longer be maintained, the system reduces its potential energy by forming a transient polaron. Here $\gamma_{\mathrm{min}}$ jumps to $1$ and the effective dot level is lowered by the full polaron binding energy $\varepsilon_p$. The spectral functions in the vicinity of this transition show that the spectral weight is redistributed to inelastic channels within the Fermi window. Consequently, the current shows no discontinuity or NDC at this point.
%
%
%
%
\section{Summary}\label{SECsummary}
In this work, we investigate the steady-state transport through a vibrating molecular quantum dot. Within the Kadanoff-Baym formalism, the nonequilibrium dot self-energy is calculated to second order in the interaction coefficients. To describe the polaronic character of the quantum dot state, we apply a variational Lang-Firsov transformation and determine the degree of transformation self-consistently by minimizing the thermodynamic potential. 

In this framework we are able to study the molecular junction for all ratios of the dot-lead coupling to the energy of the local phonon mode, i.e. from the adiabatic to the antiadiabatic regime.  Moreover, the EP interaction can be varied from weak to strong coupling. Tuning the electronic dot level and the external voltage bias, we can finally consider resonant and off-resonant transport in the equilibrium and nonequilibrium situation.

In the adiabatic regime, we find important corrections to the result of the SCBA when the EP coupling grows: In the equilibrium, off-resonant situation, the mean-field oscillator shift is reduced and spectral weight is transferred from elastic to inelastic channels. For finite voltages, we observe a pronounced peak in the electron tunneling signal, followed by several pronounced multiphonon features.  

In the antiadiabatic regime, away from the very strong coupling limit, the weight of the transient polaron state is smaller than predicted by the complete Lang-Firsov transformation. Accordingly, the equilibrium linear conductance as well as the low voltage resonant tunneling current increase, because few-phonon emission processes are amplified. As the voltage bias grows the full Lang-Firsov polaron forms. Here, due to a dynamical renormalization of the dot-lead coupling, we find NDC along the resonant current plateaus.

Most notably, our variational approach also allows the investigation of the intermediate regime where the dot-lead coupling and the phonon energy are of the same order. For weak EP coupling, the linear conductance shows a single resonance peak as a function of the electronic dot level. When the coupling strength is increased this peak narrows and shifts, signaling the crossover from coherent tunneling to sequential hopping via a long-living, transient polaron at the dot. For very strong coupling, the polaron formation takes place discontinuously, as the system switches between various metastable states.
At finite temperatures, this transition becomes continuous again. At the same time, the equilibrium linear conductance signal broadens and shows distinct phonon side peaks. Thermally activated transport via phonon absorption induces polaron formation far from resonance. 
In the low-temperature, nonequilibrium situation, the differential conductance remains positive for all voltages: the polaron induced renormalization of the dot-lead coupling is compensated by the onset of off-resonant inelastic transport. In the off-resonant, high-voltage regime, the polaron level follows the lead chemical potential to enhance resonant transport and maximize the kinetic energy.
 
Let us emphasize that we determine the current through the dot by means of an approximation to the electronic spectral function that contains inelastic features to all orders in the EP coupling. We compare our results to an effective electron model, which accounts for the electron-phonon interaction only via a renormalized dot-lead coupling parameter (e.g. in analogy to Ref. \onlinecite{LD07}). For this model negative differential conductance is observed. This is because the effective electronic spectral function does not include inelastic features that affect transport for voltages exceeding the phonon frequency.

The present study may be extended in several directions: ($\mathrm{i}$) description of hysteretic behavior in the strong coupling, high voltage regime; ($\mathrm{ii}$) inclusion of the dynamics of the phonon subsystem by means of nonequilibrium phonon Green functions; ($\mathrm{iii}$) incorporation of Coulomb interaction at the dot to produce even stronger nonlinear effects through the competition of a population-dependent repulsive dot potential with the polaronic level shift.

\begin{acknowledgments}

This work was supported by Deutsche Forschungsgemeinschaft through SFB 652 B5. TK and HF acknowledge the hospitality at the Institute of Physics ASCR.

\end{acknowledgments}
%
%
%
\section*{Appendix: Derivation of the current formula}
Deducing the current response in Sec.~\ref{SECcurrent}, the following real-time Green functions (defined according to Mahan \cite{Mah00}) are used:
\begin{align}
g^t(t_1,t_2)&=\Theta(t_1-t_2)g^>(t_1,t_2) \label{EQUtimeordered}\\
&\quad+\Theta(t_2-t_1)g^<(t_1,t_2)\;,\nonumber\\
g^{\bar t} (t_1,t_2)&=\Theta(t_2-t_1)g^>(t_1,t_2)\label{EQUantitimeordered}\\
&\quad+\Theta(t_1-t_2)g^<(t_1,t_2)\;,\nonumber
\end{align}
where $\Theta$ is the Heaviside function. The relations of $g^t$ and $g^{\bar t}$ to the retarded and advanced Green functions read
\begin{align}
g^{\mathrm{ret}}=g^{t}-g^{<}=g^{>}-g^{\bar t}\label{EQUretarded} \;,\\
g^{\mathrm{adv}}=g^{t}-g^{>}=g^{<}-g^{\bar t}\label{EQUadvanced}\;,
\end{align}
and Eq.~(\ref{EQUrealeom1}) may be written as
\begin{align}
-\frac{\sqrt{N}}{t_{ka}^\ast} \widetilde g_{cd}^< & (k,a;t_1,t_2;U)=  \label{EQUrealeom2}\\
&\quad\int_{-\infty}^{\infty}\mathrm{d}\bar t_1\; g_{cc}^{(0)t}(k,a;t_1,\bar t_1;U) \widetilde g_{dd}^<(\bar t_1,t_2;U)\nonumber\\
&- \int_{-\infty}^{\infty}\mathrm{d}\bar t_1\; g_{cc}^{(0)<}(k,a;t_1,\bar t_1;U) \widetilde g_{dd}^{\bar t}(\bar t_1,t_2;U)\nonumber\;.
\end{align}
As far as the steady-state is concerned, all averages in the definitions of the Green functions above dependent only on the differences of time variables. Consequently, the integrals on the right-hand side of Eq.~(\ref{EQUrealeom2}) may be rewritten in the form of a convolution and the Fourier transformation of Eq.~(\ref{EQUrealeom2}) is
\begin{align}
\widetilde g_{cd}^<(k,a;\omega;U) &= -\frac{t_{ka}^\ast}{\sqrt{N}} \Big [ g_{cc}^{(0)t}(k,a;\omega;U)\widetilde g_{dd}^<(\omega;U) \label{EQUrealft} \\
&\quad-g_{cc}^{(0)<}(k,a;\omega;U)\widetilde g_{dd}^{\bar t}(\omega;U)\Big]\;.\nonumber
\end{align}
Here, the Fourier transforms of the response functions are defined in the usual convention, i.e. without the factors $\pm\mathrm{i}$ introduced by Eqs.~(\ref{EQUtrafoom}) and (\ref{EQUtrafot}). In particular, the conventional Fourier transforms fulfil
\begin{equation}
\left [g^{\lessgtr}(\omega)\right]^\ast=-g^{\lessgtr}(\omega)\label{EQUglessgtrconj}\;,
\end{equation} 
because the left-hand side of Eq.~(\ref{EQUtrafoom}) is a real function. Taking into account the general property that $[g^{\mathrm{ret}}(\omega)]^\ast=g^{\mathrm{adv}}(\omega)$, the relations (\ref{EQUretarded}), (\ref{EQUadvanced}) and (\ref{EQUglessgtrconj}) give
\begin{equation}
\left [g^{t}(\omega)\right]^\ast=-g^{\bar t}(\omega)\label{EQUgtconj}\;.
\end{equation} 
With the help of Eqs.~(\ref{EQUglessgtrconj}) and (\ref{EQUgtconj}), the complex conjugate of $\widetilde g_{cd}^{\lessgtr}(k,a;\omega;U)$ in Eq.~(\ref{EQUmeancd}) is determined and the following formula for the current $J_a$ results:
\begin{widetext}
\begin{align}\label{EQUcurrenta1}
J_a&= -\frac{\mathrm{e}}{N} \sum_k |t_{ka}|^2 \int_{-\infty}^{\infty} \frac{\mathrm{d}\omega}{2\pi}\; \Big \{ \; 
\Big [ g_{cc}^{(0)t}(k,a;\omega;U) + g_{cc}^{(0)\bar t}(k,a;\omega;U) \Big ] \widetilde g_{dd}^{<}(\omega;U) \\
&\quad-g_{cc}^{(0)<}(k,a;\omega;U) \Big [ \widetilde g_{dd}^{t}(\omega;U) + \widetilde g_{dd}^{\bar t}(\omega;U) \Big ] \; \Big \}\;.\nonumber
\end{align}
Substituting the explicit forms of the free electron functions $g_{cc}^{(0)\lessgtr}$ and using the relation $g^t+g^{\bar t}=g^>+g^<$ following from Eqs.~(\ref{EQUretarded}) and (\ref{EQUadvanced}), we obtain
\begin{align}\label{EQUcurrenta2}
J_a&=-\frac{\mathrm{e}}{N} \sum_k |t_{ka}|^2 \int_{-\infty}^{\infty} \frac{\mathrm{d}\omega}{2\pi}\; 2\pi\delta(\omega-\xi_{ka})\Big \{ -\mathrm{i} \widetilde g_{dd}^{<}(\omega;U) + f(\xi_{ka}+U_a) \mathrm{i} \Big [ \widetilde g_{dd}^<(\omega;U)-\widetilde g_{dd}^>(\omega;U) \Big ]  \Big \} \;.
\end{align}
Going back to the definitions of the Fourier transforms according to Eqs.~(\ref{EQUtrafoom}) and (\ref{EQUtrafot}), we arrive at Eq.~(\ref{EQUcurrenta}) of Sec.~\ref{SECcurrent}.
\end{widetext}
%
%


%
\end{document}